\documentclass[pof,twocolumn,amsmath,amssymb,reprint,nofootinbib]{revtex4-2}
\usepackage{epsfig}                                                                 % Include eps figure files
\usepackage{dcolumn}                                                                % Align table columns on decimal point
\usepackage{bm}                                                                     % Bold math

\PassOptionsToPackage{hyphens}{url}
\usepackage[pdfstartview=FitH,bookmarksopen=true,bookmarksopenlevel=2]{hyperref}    % Include hypertext links with bookmarks
\usepackage{url}

\begin{document}
\title[Nonisothermal evaporation]{Nonisothermal evaporation}
\author{E. S. Benilov}
 \email[Email address: ]{Eugene.Benilov@ul.ie}
 \homepage[\newline Homepage: ]{https://staff.ul.ie/eugenebenilov/}
 \affiliation{Department of Mathematics and Statistics, University of Limerick, Limerick V94~T9PX, Ireland}

\begin{abstract}
Evaporation of a liquid layer on a substrate is examined without the
often-used isothermality assumption -- i.e., temperature variations are
accounted for. Qualitative estimates show that nonisothermality makes the
evaporation rate depend on the conditions the substrate is maintained at. If
it is thermally insulated, evaporative cooling dramatically slows evaporation
down; the evaporation rate tends to zero with time and cannot be determined by
measuring the external parameters only. If, however, the substrate is
maintained at a fixed temperature, the heat flux coming from below sustains
evaporation at a finite rate -- deducible from the fluid's characteristics,
relative humidity, and the layer's depth (whose importance has not been
recognized before). The qualitative predictions are quantified using the
diffuse-interface model applied to a liquid evaporating into its own vapor.

\end{abstract}
\maketitle

\section{Introduction and preliminary estimates}

Evaporation of liquids has been studied for over a century, since the
pioneering work of James Clerk Maxwell \cite{Maxwell77,Maxwell90}, and it is
still studied now -- both theoretically (e.g., Refs.
\cite{SobacTalbotHautRednikovColinet15,TalbotSobacRednikovColinetHaut16,Sazhin17,WilliamsKarapetsasMamalisSefianeMatarValluri20,FinneranGarnerNadal21,DallabarbaWangPicano21}%
) and experimentally (e.g., Refs.
\cite{JakubczykKolwasDerkachovKolwasZientara12,HolystLitniewskiJakubczyk17,SchweiglerBensaidSeifritzSelzerNestler17,PoosVarju20,VarjuPoos22}%
) -- as numerous issues have yet to be resolved.

Consider, for example, a flat liquid layer. It is generally believed that it
evaporates at a steady rate depending on the liquid's parameters (temperature,
heat of vaporization, etc.) and the humidity of air. There are numerous
measurements of evaporation rates; a recent review of this work in application
to water can be found in Refs. \cite{PoosVarju20,VarjuPoos22}.

\begin{figure}
\includegraphics[width=\columnwidth]{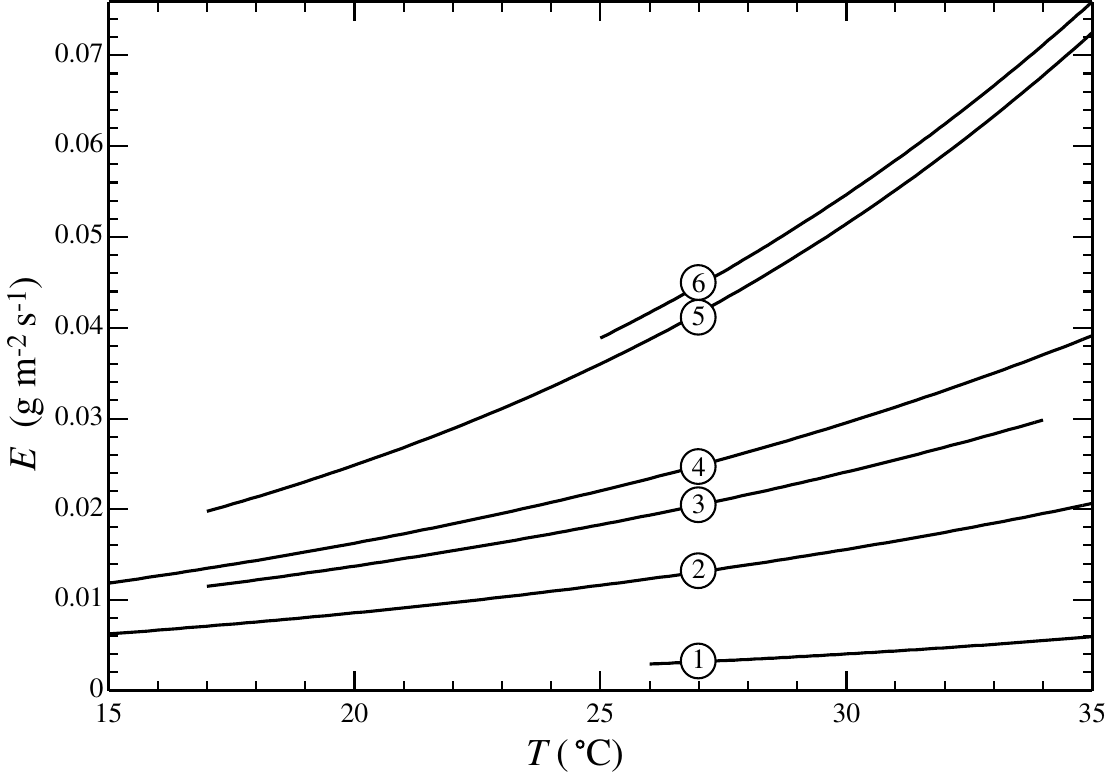}
\caption{Evaporation rate vs. temperature, according to various empiric formulae. Curves (1)--(6) correspond to Refs. \cite{MoghimanJodat07,MackayVanwesenbeeck14,SharpleyBoelter38,Shah12,BoelterGordonGriffin46,TangPaukenJeterAbdelkhalik93} , respectively. In all cases, the relative humidity is 50\%.}
\label{fig1}
\end{figure}

Fig. \ref{fig1} shows a selection\footnote{Fig. \ref{fig1} shows those empiric
formulae listed in Table 1 of Ref. \cite{VarjuPoos22} that involve only the
relative humidity and characteristics uniquely related to the temperature,
such as the saturated pressure and vaporization heat of water (calculated
using Refs. \cite{WagnerPruss02,Hendersonsellers84}, respectively). One of the
formulae includes also the wind speed which was set to zero.
\par
All the other empiric results cited in Ref. \cite{VarjuPoos22} involve further
parameters (e.g., the horizontal scale\ of the vessel), making a comparison
with the low-parameter formulae impossible.} of measurements of the
evaporation rate $E$, for water evaporating into still air, as a function of
the temperature $T$ within a \textquotedblleft room
temperature\textquotedblright\ range. There is evident discord in these
results, suggesting that important factors vary from experiment to experiment.

The present paper identifies at least some of these factors. It is shown that
nonisothermal effects -- e.g., the heat exchange between the liquid and
substrate (and side walls, if any) -- can make $E$ depend on the distance
between the interface and substrate, and the material the latter is made of.

To illustrate the importance of heat exchange with the boundaries, consider an
amount of liquid in a thermally insulated vessel -- and let half of the liquid
evaporate. The temperature of the remaining half decreases due to evaporative
cooling -- and the size of the decrease is easy to estimate. Assuming for
simplicity that vaporization heat $\Delta h$ and heat capacity $c_{P}^{(l)}$
of the liquid do not change significantly with $T$, one can approximate the
temperature decrease by%
\begin{equation}
\Delta T=\frac{\Delta h}{c_{P}^{(l)}}. \label{1.1}%
\end{equation}
For $\Delta h=2442\,\mathrm{J\,g}^{-1}$ and $c_{P}^{(l)}=4.182\,\mathrm{J\,g}%
^{-1}\mathrm{K}^{-1}$ (which correspond to water at $25^{\circ}\mathrm{C}$
\cite{LindstromMallard97}), Eq. (\ref{1.1}) yields a somewhat unexpected
result:%
\[
\Delta T\approx584\,\mathrm{K}.
\]
In reality, however, evaporation of liquid in an insulated vessel slows down
to a virtual standstill well before it half-evaporates. Since the dependence
of $E$ on $T$ at normal conditions is typically exponential, even a moderate
temperature decrease can reduce the evaporation rate by an order of magnitude.

Alternatively, let the vessel's walls and bottom be kept at a fixed
temperature (such a setting has probably more applications). In this case, the
energy loss to vaporization is replenished by the incoming heat flux, which
can be readily calculated,%
\begin{equation}
-\kappa^{(l)}T^{\prime}=E\,\Delta h, \label{1.2}%
\end{equation}
where $\kappa^{(l)}$ is the liquid's thermal conductivity and the temperature
gradient $T^{\prime}$ can be expressed through the temperature difference
$\Delta T$ between the interface and the nearest boundary, and the
corresponding distance $D$,%
\begin{equation}
T^{\prime}=-\frac{\Delta T}{D}. \label{1.3}%
\end{equation}
To determine $\Delta T$ for water at $25^{\circ}\mathrm{C}$, set $\kappa
^{(l)}=0.6065\,\mathrm{W\,m}^{-1}\mathrm{K}^{-1}$ \cite{LindstromMallard97}
and%
\begin{equation}
E=0.025361\,\mathrm{g\,m}^{-2}\mathrm{s}^{-1}, \label{1.4}%
\end{equation}
which is the average of empiric curves 2--6 in Fig. \ref{fig1} (curve 1 cannot
be used, as $25^{\circ}\mathrm{C}$ is not in its range). With these values,
Eqs. (\ref{1.2})--(\ref{1.3}) yield%
\begin{equation}
\Delta T=\frac{E\,\Delta h}{\kappa^{(l)}}D\approx\frac{1\,\mathrm{K}%
}{1\,\mathrm{cm}}\times D. \label{1.5}%
\end{equation}
Evidently, this estimate is both qualitatively and quantitatively different
from that for insulated substrates.

The difference between insulated and fixed-$T$ vessels demonstrates the
importance of heat fluxes from boundaries and, generally, nonisothermal
effects. In this work, they are explored using the simplest setting:
evaporation of a liquid into \emph{its own undersaturated vapor}. It is
described by a relatively simple model which does not include the diffusive
mass flux (pure fluids do not diffuse). Evaporation in this case occurs via
advection \cite{Benilov22a}, but heat conduction is similar to that in mixtures.

The described setting will be examined using the so-called diffuse interface
model (DIM). It was proposed in 1901 by Diederik Korteweg \cite{Korteweg01}
and has been used since then in thousands of papers and for tens of
applications (some of this work is reviewed in Ref. \cite{Benilov23a}). The DIM
is particularly suited to the problem at hand: it describes both liquid and
vapor, as well as the interfacial dynamics -- as opposed to models built of
`blocks' describing one item each. The use of the DIM is convenient but not
crucial, however, as nonisothermal effects can be introduced into any good
model of evaporation.

In Sec. \ref{Sec 2} of the present paper, the problem is formulated
mathematically. In Sec. \ref{Sec 3}, evaporation is examined under the
assumption of isothermality. This case will be used as a yardstick for the
full problem examined in \ref{Sec 4}. Other effects potentially explaining the
discord among the empiric curves in Fig. \ref{fig1} are discussed in Sec.
\ref{Sec 5}. In Sec. \ref{Sec 6}, the results are summarized, plus it is
clarified there when the heat flux from air is weak and its effect on
evaporation, negligible (which is of interest in a broader context, not just
for the present work).

\section{Formulation\label{Sec 2}}

\subsection{Thermodynamics\label{Sec 2.1}}

Thermodynamic properties of a fluid can be described by the dependence of its
specific (per unit mass) internal energy $e$ and specific entropy $s$ on the
density $\rho$ and temperature $T$. The functions $e(\rho,t)$ and $s(\rho,t)$
are supposed to be constrained by the Gibbs relation; if written in terms of
$e$ and $s$, it takes the form%
\begin{equation}
\frac{\partial e}{\partial T}=T\frac{\partial s}{\partial T}. \label{2.1}%
\end{equation}
The fluid's equation of state, or the expression for the pressure $p$, is
given by%
\begin{equation}
p=\rho^{2}\left(  \frac{\partial e}{\partial\rho}-T\frac{\partial s}%
{\partial\rho}\right)  , \label{2.2}%
\end{equation}
and the chemical potential, or specific free energy, by%
\begin{equation}
G=e+\frac{p}{\rho}-Ts. \label{2.3}%
\end{equation}
It follows from (\ref{2.1})--(\ref{2.3}) that%
\begin{equation}
\frac{\partial p}{\partial\rho}=\rho\dfrac{\partial G}{\partial\rho},
\label{2.4}%
\end{equation}%
\begin{equation}
\dfrac{\partial p}{\partial T}=-\rho^{2}\frac{\partial s}{\partial\rho}.
\label{2.5}%
\end{equation}
These two identities enable one to replace $p$ with $G$ or $s$, which happens
to be convenient in the problem at hand.

Define the heat capacity at constant volume,%
\begin{equation}
c_{V}=\frac{\partial e}{\partial T}, \label{2.6}%
\end{equation}
the specific enthalpy,%
\begin{equation}
h=e+\frac{p}{\rho}, \label{2.7}%
\end{equation}
and the heat capacity at constant pressure,%
\[
c_{P}=\left(  \frac{\partial h}{\partial T}\right)  _{p=\operatorname{const}%
}.
\]
Expressing the derivative at constant pressure via the partial derivatives
with respect $\rho$ and $T$, and recalling Eq. (\ref{2.7}), one obtains%
\begin{equation}
c_{P}=\frac{\partial e}{\partial T}-\left(  \dfrac{\partial e}{\partial\rho
}-\frac{p}{\rho^{2}}\right)  \dfrac{\partial p}{\partial T}\left(
\dfrac{\partial p}{\partial\rho}\right)  ^{-1}. \label{2.8}%
\end{equation}

\subsection{Hydrodynamics\label{Sec 2.2}}

Consider a liquid layer on a horizontal solid substrate, and vapor above the
liquid. If the vapor in undersaturated, the liquid evaporates, giving rise to
a vertical flow. This setting is characterized by the velocity $w(z,t)$,
density $\rho(z,t)$, and temperature $T(z,t)$, where $z$ is the vertical
coordinate and $t$, the time.

\subsubsection{Governing equations\label{Sec 2.2.1}}

It can be safely assumed that the Reynolds number associated with evaporation
is small, so that the contribution of inertia to the balance of momentum is
negligible. Thus, Stokes-flow (slow-flow) approximation can be employed.

Using the diffuse-interface model (DIM), one can write the Stokes-flow version
of the hydrodynamic equations in the form%
\begin{equation}
\frac{\partial\rho}{\partial t}+\frac{\partial}{\partial z}%
\underset{\text{mass flux}}{\underbrace{\left(  \rho w\right)  }}=0,
\label{2.9}%
\end{equation}%
\begin{equation}
\underset{\text{pressure gradient}}{\underbrace{\frac{\partial p}{\partial z}%
}}~=~\underset{\text{viscous stress}}{\underbrace{\frac{\partial}{\partial
z}\left(  \eta\frac{\partial w}{\partial z}\right)  }}~~+\underset{\text{van
der Waals force}}{\underbrace{K\rho\frac{\partial^{3}\rho}{\partial z^{3}}}},
\label{2.10}%
\end{equation}%
\begin{multline}
\frac{\partial\left(  \rho e\right)  }{\partial t}+\frac{\partial}{\partial
z}\underset{\text{heat flux}}{\underbrace{\left[  w\left(  \rho e+p-\eta
\frac{\partial w}{\partial z}\right)  -\kappa\frac{\partial T}{\partial
z}\right]  }}\\
=\underset{\text{rate of work done by van der Waals force}%
}{\underbrace{w\,K\rho\frac{\partial^{3}\rho}{\partial z^{3}}}}, \label{2.11}%
\end{multline}
where the effective viscosity $\eta$ is related to the shear viscosity
$\mu_{s}$ and and bulk viscosity $\mu_{b}$ by%
\[
\eta=\frac{4}{3}\mu_{s}+\mu_{b},
\]
$\kappa$ is the thermal conductivity and $K$, the so-called Korteweg
parameter. The expression for the van der Waals force comes from the DIM,
otherwise (\ref{2.9})--(\ref{2.11}) are standard equations of compressible
Stokes-flow hydrodynamics. The three-dimensional, non-Stokes flow versions of
Eqs. (\ref{2.9})--(\ref{2.11}) were derived in Ref. \cite{Giovangigli20} from
the Enskog--Vlasov kinetic theory, and in Ref. \cite{GalloMagalettiCasciola21}
via nonequilibrium thermodynamics. In this paper, a brief derivation of the
DIM expression for the van der Waals force is given in Appendix
\ref{Appendix A}.

Note that $\eta$ and $\kappa$ depend generally on $\rho$ and $T$, whereas $K$
is a constant. Its value is related to, and can be deduced from, the fluid's
surface tension -- for water, for example, $K\approx1.9\times10^{-17}%
\mathrm{m}^{7}\mathrm{s}^{-2}\mathrm{kg}^{-1}$ \cite{Benilov23a}.

\subsubsection{Boundary conditions far above the interface\label{Sec 2.2.3}}

Assume that, far above the liquid--vapor interface, the viscous stress is
zero,%
\begin{equation}
\frac{\partial w}{\partial z}\rightarrow0\qquad\text{as}\qquad z\rightarrow
+\infty, \label{2.12}%
\end{equation}
and the vapor density and temperature tend to certain values,%
\begin{equation}
T\rightarrow T^{(v)}\qquad\text{as}\qquad z\rightarrow+\infty, \label{2.13}%
\end{equation}%
\begin{equation}
\rho\rightarrow\rho^{(v)}\qquad\text{as}\qquad z\rightarrow+\infty.
\label{2.14}%
\end{equation}
For evaporation to occur, $\rho^{(v)}$ should be smaller than the saturated
vapor density $\rho^{(v.sat)}$ -- which is determined, together with the
matching saturated liquid density $\rho^{(l.sat)}$, by the Maxwell
construction:%
\begin{equation}
p(\rho^{(l.sat)},T^{(v)})=p(\rho^{(v.sat)},T^{(v)}), \label{2.15}%
\end{equation}%
\begin{equation}
G(\rho^{(l.sat)},T^{(v)})=G(\rho^{(v.sat)},T^{(v)}), \label{2.16}%
\end{equation}
where it is implied that $\rho^{(l.sat)}\geq\rho^{(v.sat)}$.

Physically, the equalities of the pressure and chemical potential in the two
phases guarantee the mechanical and thermodynamic equilibria of the interface,
respectively. Mathematically, conditions (\ref{2.15})--(\ref{2.16}) can be
derived from the DIM (to be elaborated later) or from any other good model
describing a static flat interface in an unbounded space.

Require also that the saturated vapor and liquid be thermodynamically stable,
which amounts to%
\begin{equation}
\left(  \frac{\partial p}{\partial\rho}\right)  _{\rho=\rho^{(v.sat)}}%
\geq0,\qquad\left(  \frac{\partial p}{\partial\rho}\right)  _{\rho
=\rho^{(l.sat)}}\geq0, \label{2.17}%
\end{equation}
i.e., an increase in $\rho$ does not decrease the pressure. The Maxwell
construction (\ref{2.15})--(\ref{2.16}) and requirements (\ref{2.17}) uniquely
define $\rho^{(v.sat)}$ and $\rho^{(l.sat)}$ as functions of $T$.

Before proceeding further, it is convenient to integrate the momentum equation
(\ref{2.10}) and fix the constant of integration via boundary conditions
(\ref{2.12})--(\ref{2.14}), which yields%
\begin{equation}
p=\eta\frac{\partial w}{\partial z}+K\left[  \rho\frac{\partial^{2}\rho
}{\partial z^{2}}-\frac{1}{2}\left(  \frac{\partial\rho}{\partial z}\right)
^{2}\right]  +p^{(v)}, \label{2.18}%
\end{equation}
where $p^{(v)}=p(\rho^{(v)},T^{(v)})$.

\subsubsection{Boundary conditions at the substrate\label{Sec 2.2.2}}

Let the substrate be located at $z=0$ and write the no-flow boundary condition
in the form%
\begin{equation}
w=0\qquad\text{at}\qquad z=0. \label{2.19}%
\end{equation}
Two different conditions for $T$ will be examined: one assuming that the
substrate is kept at a fixed temperature,%
\begin{equation}
T=T_{0}\qquad\text{at}\qquad z=0, \label{2.20}%
\end{equation}
and another corresponding to a thermally insulated substrate,%
\begin{equation}
\frac{\partial T}{\partial z}=0\qquad\text{at}\qquad z=0. \label{2.21}%
\end{equation}
Due to the presence of higher-order derivatives of $\rho$ in expression for
the van der Waals force, a separate boundary condition is required for the
density. Several different versions of such are used in the literature (e.g.,
Refs. \cite{Seppecher96,PismenPomeau00,GalloMagalettiCasciola21}), describing
slightly different models of the fluid--substrate interaction. The specific
form of this boundary condition is important only if the thickness of the
liquid--vapor interface is comparable to the liquid layer's depth -- which is,
obviously, not the case for a \emph{macroscopic} layer considered in this work.

Thus, the simplest version of the boundary condition for $\rho$ will be used
-- the one suggested in Ref. \cite{PismenPomeau00},%
\begin{equation}
\rho=\rho_{0}\qquad\text{at}\qquad z=0, \label{2.22}%
\end{equation}
where $\rho_{0}$ characterizes the fluid--substrate interaction. This
condition reflects the balance of forces affecting fluid molecules adjacent to
the substrate, and it can be derived under the same assumptions as the DIM
itself \cite{Benilov20a}.

Note that none of the conclusions reported in this paper depends on the
specific value of $\rho_{0}$.

\subsubsection{A flat interface in an unbounded fluid\label{Sec 2.2.4}}

In a sufficiently deep layer, the interface is not affected by the substrate.
Boundary condition (\ref{2.19}) can be moved to minus-infinity,%
\begin{equation}
w\rightarrow0\qquad\text{as}\qquad z\rightarrow-\infty, \label{2.23}%
\end{equation}
and conditions (\ref{2.20})--(\ref{2.22}), replaced with%
\begin{equation}
T\rightarrow T^{(l)}\qquad\text{as}\qquad z\rightarrow-\infty, \label{2.24}%
\end{equation}%
\begin{equation}
\rho\rightarrow\rho^{(l)}\qquad\text{as}\qquad z\rightarrow-\infty.
\label{2.25}%
\end{equation}
Note that Eq. (\ref{2.18}) and boundary conditions (\ref{2.23})--(\ref{2.25})
imply that the liquid's temperature $T^{(l)}$ and density $\rho^{(l)}$ are not
entirely arbitrary, but are related to the vapor parameters by%
\begin{equation}
p(\rho^{(l)},T^{(l)})=p(\rho^{(v)},T^{(v)}). \label{2.26}%
\end{equation}
Mathematically, this constraint is a result of the Stokes-flow approximation:
if the time derivative in the momentum equation were retained, (\ref{2.26})
would not hold. Physically, the constraint suggests that an adjustment of
pressure should occur (via fast acoustic waves) before the flow becomes truly slow.

\subsection{Nondimensionalization\label{Sec 2.3}}

To nondimensionalize Eqs. (\ref{2.9})--(\ref{2.11}), introduce characteristic
scales of pressure $P$, density $\varrho$, and viscosity $\bar{\eta}$. Using
these parameters, one can define a velocity scale and a spatial scale,%
\[
V=\frac{Pl}{\bar{\eta}},\qquad l=\sqrt{\frac{K\varrho^{2}}{P}}.
\]
As seen later, this choice of $V$ and $l$ corresponds to an asymptotic regime
where the pressure gradient, viscous stress, and van der Waals force in Eq.
(\ref{2.10}) are all of the same order. The interfacial scale $l$ will be
referred to as `microscopic', and the depth of the whole liquid layer, `macroscopic'.

The following nondimensional variables will be used:%
\[
z_{nd}=\frac{z}{l},\qquad t_{nd}=\frac{V}{l}t,
\]%
\[
w_{nd}=\frac{w}{V},\qquad\rho_{nd}=\frac{\rho}{\varrho},\qquad T_{nd}%
=\frac{\varrho RT}{P},
\]%
\[
e_{nd}=\frac{\varrho e}{P},\qquad s_{nd}=\frac{s}{R},
\]%
\[
p_{nd}=\frac{p}{P},\qquad G_{nd}=\frac{\varrho G}{P},\qquad h_{nd}%
=\frac{\varrho h}{P},
\]%
\[
\left(  c_{V}\right)  _{nd}=\frac{c_{V}}{R},\qquad\left(  c_{P}\right)
_{nd}=\frac{c_{P}}{R},
\]
where $R$ is the specific gas constant of the fluid under consideration. It is
convenient to also nondimensionalize the viscosity and thermal diffusivity,%
\[
\eta_{nd}=\frac{\eta}{\bar{\eta}},\qquad\kappa_{nd}=\frac{\kappa}{\bar{\kappa
}},
\]
where $\bar{\eta}$ is a characteristic value of $\eta$, whereas $\bar{\kappa}$
is \emph{not} that of $\kappa$ -- but is given by%
\[
\bar{\kappa}=\frac{KR\varrho^{3}}{\bar{\eta}}.
\]
This choice of $\bar{\kappa}$ conveniently eliminates all nondimensional
parameters in the governing equations, but one should keep in mind that
$\kappa_{nd}$ can be large or small.

In terms of the new variables, Eqs. (\ref{2.9})--(\ref{2.11}) take the form
(the subscript $_{nd}$ omitted)%
\begin{equation}
\frac{\partial\rho}{\partial t}+\frac{\partial\left(  \rho w\right)
}{\partial z}=0, \label{2.27}%
\end{equation}%
\begin{equation}
p=\eta\frac{\partial w}{\partial z}+\rho\frac{\partial^{2}\rho}{\partial
z^{2}}-\frac{1}{2}\left(  \frac{\partial\rho}{\partial z}\right)  ^{2}%
+p^{(v)}, \label{2.28}%
\end{equation}%
\begin{multline}
\frac{\partial\left(  \rho e\right)  }{\partial t}+\frac{\partial}{\partial
z}\left[  w\left(  \rho e+p-\eta\frac{\partial w}{\partial z}\right)
-\kappa\frac{\partial T}{\partial z}\right] \\
=w\rho\frac{\partial^{3}\rho}{\partial z^{3}}. \label{2.29}%
\end{multline}
The nondimensional versions of the thermodynamics identities from Sec.
\ref{Sec 2.1} look exactly as their dimensional counterparts, and so do the
boundary conditions from Secs. \ref{Sec 2.2.2}--\ref{Sec 2.2.3}, provided
$\rho_{0}$ is nondimensionalized by $\varrho$ and $T_{0}$ by $P/\varrho R$.

\subsection{The van der Waals fluid\label{Sec 2.4}}

\begin{figure}
\includegraphics[width=\columnwidth]{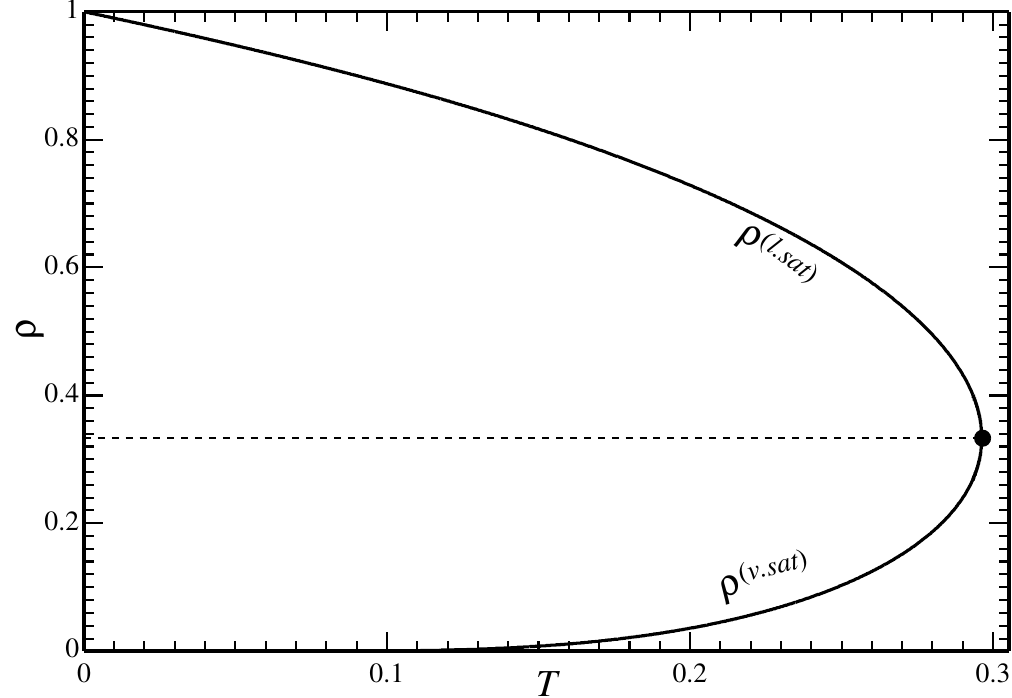}
\caption{The saturated densities of liquid and vapor vs. the temperature, for the van der Waals fluid. The nondimensional critical point in this case is $T_{cr}=8/27$ (marked by the black dot and dotted line).}
\label{fig2}
\end{figure}

In what follows, general conclusions will be illustrated using the van der
Waals fluid, whose nondimensional internal energy and entropy are%
\begin{equation}
e=c_{V}T-\rho,\qquad s=c_{V}\ln T-\ln\frac{\rho}{1-\rho}, \label{2.30}%
\end{equation}
where the heat capacity at constant volume $c_{V}$ is a given constant. Then,
Eqs. (\ref{2.2})--(\ref{2.3}) yield the following expressions for the pressure
and chemical potential:%
\begin{equation}
p(\rho,T)=\frac{T\rho}{1-\rho}-\rho^{2}, \label{2.31}%
\end{equation}%
\begin{multline}
G(\rho,T)=T\left(  \ln\frac{\rho}{1-\rho}+\frac{1}{1-\rho}\right)  -2\rho\\
+c_{V}T\left(  1-\ln T\right)  . \label{2.32}%
\end{multline}
To illustrate the properties of the van der Waals fluid, expressions
(\ref{2.31})--(\ref{2.32}) and the Maxwell construction (\ref{2.15}%
)--(\ref{2.16}) were used to compute the saturated densities $\rho^{(l.sat)}$
and $\rho^{(v.sat)}$. The results are shown in Fig. \ref{fig2}. Observe that,
if $T>T_{cr}$, only one phase exists, so interfaces do not.

\section{Isothermal evaporation\label{Sec 3}}

The assumption of isothermality is used in many papers on evaporation (as
quantified by the 2 million results yielded by a Google search for
\textquotedblleft isothermal\textquotedblright\ + \textquotedblleft
evaporation\textquotedblright). In terms of the present model, isothermality
corresponds to the limit $\kappa\rightarrow\infty$, in which case Eq.
(\ref{2.29}) yields $\partial T/\partial z\rightarrow0$. Assuming that $T$ is
also independent of $t$, one can treat the temperature in Eqs. (\ref{2.27}%
)--(\ref{2.28}) as a known parameter. The isothermal reduction of the full DIM
was first examined in Ref. \cite{PismenPomeau00}.

\subsection{Steady evaporation\label{Sec 3.1}}

Consider first a liquid--vapor interface in an unbounded space and assume that
it is \emph{steadily} receding due to evaporation. Its velocity is equal to
$-E/\rho^{(l)}$ where $E$ is the nondimensional evaporation rate and
$\rho^{(l)}$, the liquid's nondimensional density.

\begin{figure*}
\includegraphics[width=\textwidth]{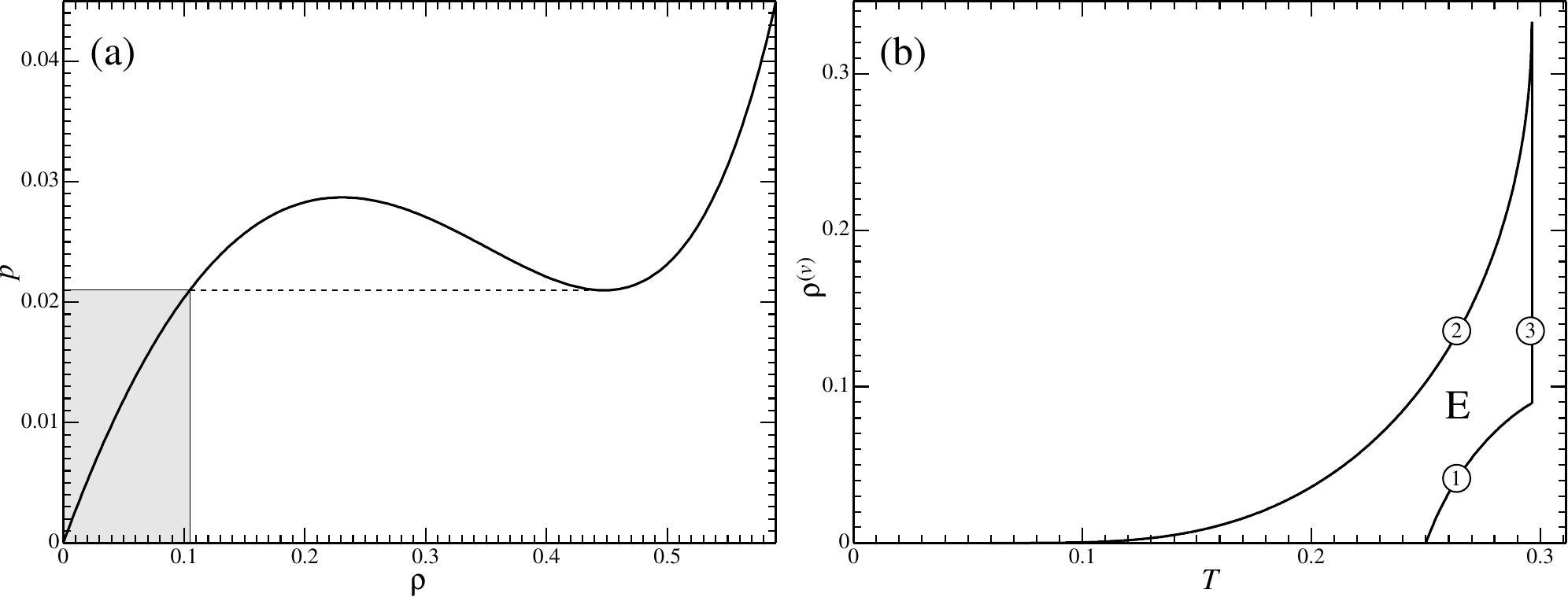}
\caption{Existence of physically meaningful solutions of Eq. (\ref{3.8}) for the van der Waals fluid. (a) The equation of state $p(\rho)$ for the particular case of $T=0.27315$; the region where the vapor density does not have a match in the liquid region is shaded. (b)~The existence region (marked with a letter \textquotedblleft E\textquotedblright) of solutions of Eq. (\ref{3.8}), in the $(T,\rho^{(v)})$ parameter plane. Curve (1) separates the values of $\rho^{(v)}$ that have a match in the liquid region from those that do not; curve 2 is the saturation curve (the vapor at infinity should be undersaturated); curve 3 is $T=T_{cr}$.}
\label{fig3}
\end{figure*}

Mathematically, the assumption of steadiness corresponds to the following
ansatz:%
\[
\rho=\rho(\xi),\qquad w=w(\xi),
\]
where%
\[
\xi=z+\frac{E}{\rho^{(l)}}t.
\]
In terms of $\xi$, the general boundary conditions (\ref{2.12}), (\ref{2.14})
and (\ref{2.23}), (\ref{2.25}) become%
\begin{equation}
w\rightarrow0\qquad\text{as}\qquad\xi\rightarrow-\infty, \label{3.1}%
\end{equation}%
\begin{equation}
\rho\rightarrow\rho^{(l)}\qquad\text{as}\qquad\xi\rightarrow-\infty,
\label{3.2}%
\end{equation}%
\begin{equation}
\frac{\mathrm{d}w}{\mathrm{d}\xi}\rightarrow0\qquad\text{as}\qquad
\xi\rightarrow+\infty, \label{3.3}%
\end{equation}%
\begin{equation}
\rho\rightarrow\rho^{(v)}\qquad\text{as}\qquad\xi\rightarrow+\infty.
\label{3.4}%
\end{equation}
When rewritten in terms of $\xi$, Eq. (\ref{2.27}) and conditions
(\ref{3.1})--(\ref{3.2}) yield%
\begin{equation}
w=E\left(  \frac{1}{\rho}-\frac{1}{\rho^{(l)}}\right)  , \label{3.5}%
\end{equation}
which automatically satisfies condition (\ref{3.3}) as well. Then Eq.
(\ref{2.28}) becomes%
\begin{multline}
p(\rho,T)=-\frac{E\eta(\rho,T)}{\rho^{2}}\frac{\mathrm{d}\rho}{\mathrm{d}\xi
}+\rho\frac{\mathrm{d}^{2}\rho}{\mathrm{d}\xi^{2}}-\frac{1}{2}\left(
\frac{\mathrm{d}\rho}{\mathrm{d}\xi}\right)  ^{2}\\
+p(\rho^{(v)},T). \label{3.6}%
\end{multline}
This equation and boundary conditions (\ref{3.2}) and (\ref{3.4}) are
translationally invariant -- thus, to ensure the solution's uniqueness, the
following extra requirement is imposed%
\begin{equation}
\rho=\frac{\rho^{(l)}+\rho^{(v)}}{2}\qquad\text{at}\qquad z=0. \label{3.7}%
\end{equation}
Eq. (\ref{3.6}) and conditions (\ref{3.2}), (\ref{3.4}), and (\ref{3.7})
constitute a boundary-value problem for the function $\rho(\xi)$ and
undetermined parameters $E$ and $\rho^{(l)}$. The latter can be found straight
away from constraint (\ref{2.26}) with $T^{(l)}=T^{(v)}=T$, which yields%
\begin{equation}
p(\rho^{(l)},T)=p(\rho^{(v)},T). \label{3.8}%
\end{equation}
This equation relates $\rho^{(l)}$ to the (known) density $\rho^{(v)}$ of the
vapor far above the interface. To make $\rho^{(l)}$ unique, one should require
that the liquid phase is stable,%
\[
\left(  \frac{\partial p}{\partial\rho}\right)  _{\rho=\rho^{(l)}}\geq0.
\]
It turns out that, if $T$ is sufficiently large and $\rho^{(v)}$ is
sufficiently small, Eq. (\ref{3.8}) does not have any solutions. Such a
situation is illustrated in Fig. \ref{fig3}a: if the vapor pressure happens to
be in the shaded region, it cannot match \emph{any} value of the liquid
pressure. Such a pattern arises only if $T>1/4$ -- otherwise the local minimum
of $p(\rho)$ is negative and, for any $\rho^{(v)}$, there exists a matching
value of $\rho^{(l)}$.

Fig. \ref{fig3}(b) shows the region of the $(T,\rho^{(v)})$ plane where Eq.
(\ref{3.8}) admits physically meaningful solutions. What happens if
$(T,\rho^{(v)})$ are such that no steady solution exists will be clarified in
Sec. \ref{Sec 2.3}.

Boundary-value problem (\ref{3.2}), (\ref{3.4}), (\ref{3.6})--(\ref{3.7}) was
solved numerically (using the function bvp5c of MATLAB) for the van der Waals
fluid (\ref{2.31}). The viscosity was assumed to be proportional to the
density, with the proportionality coefficient implied to be scaled out during
the nondimensionalization,%
\begin{equation}
\eta=\rho. \label{3.9}%
\end{equation}
This is the simplest model qualitatively reflecting growth of a fluid's
viscosity with density.

\begin{figure*}
\includegraphics[width=\textwidth]{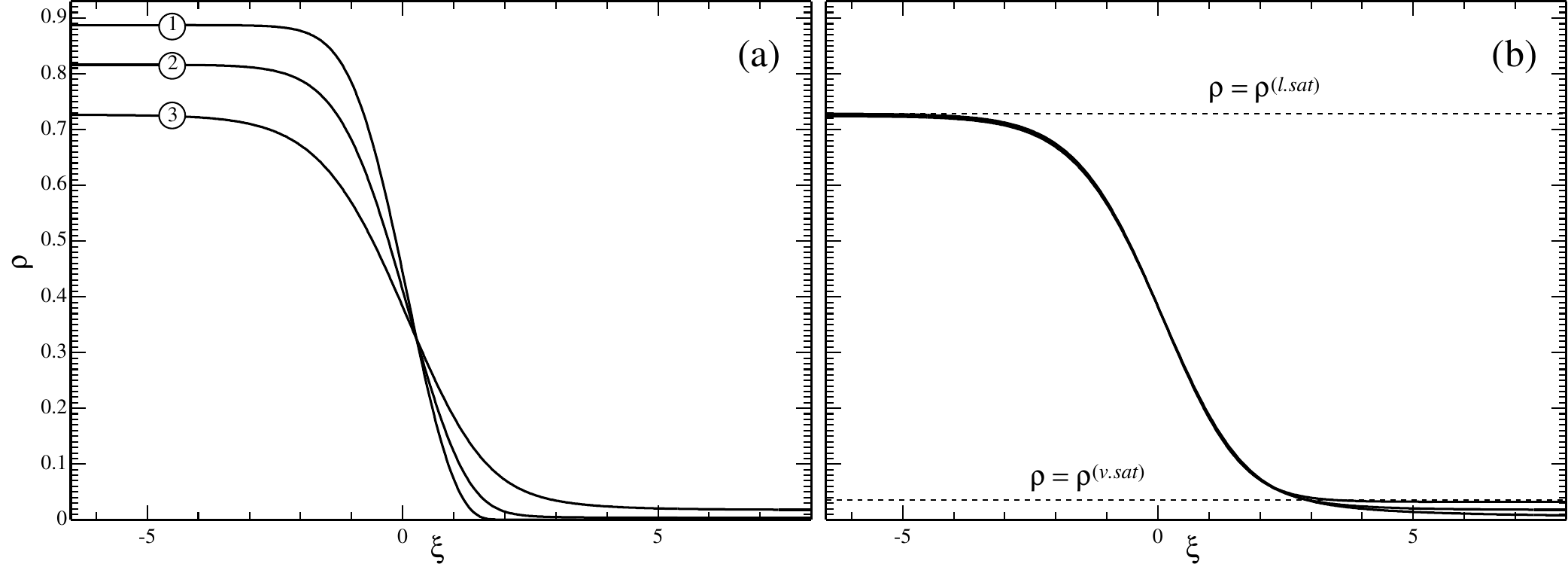}
\caption{Solutions of boundary-value problem (\ref{3.2}), (\ref{3.4}), (\ref{3.6})--(\ref{3.7}), for various temperatures and relative humidities. Panel~(a): $\rho^{(v)}/\rho^{(v.sat)}=0.5$; (1) $T=0.1$, (2) $T=0.15$, (3) $T=0.2$. Panel~(b): $T=0.2$; $\rho^{(v)}/\rho^{(v.sat)}=0.1,~0.5,~0.9$; these curves are not numbered, but can still be distinguished by comparing them to the dotted straight line marking the density $\rho^{(v.sat)}$ of saturated vapor. The other dotted line shows the density $\rho^{(l.sat)}$ of saturated liquid.}
\label{fig4}
\end{figure*}

Fig. \ref{fig4} shows typical profiles $\rho(\xi)$ for various values of $T$
[panels (a)] and various values of relative humidity $\rho^{(v)}%
/\rho^{(v.sat)}$ [panel (b)]. The latter also illustrates that, unless $T$ is
close to $T_{cr}$, the liquid's density $\rho^{(l)}$ is close to its saturated
value $\rho^{(l.sat)}$.

\begin{figure}
\includegraphics[width=\columnwidth]{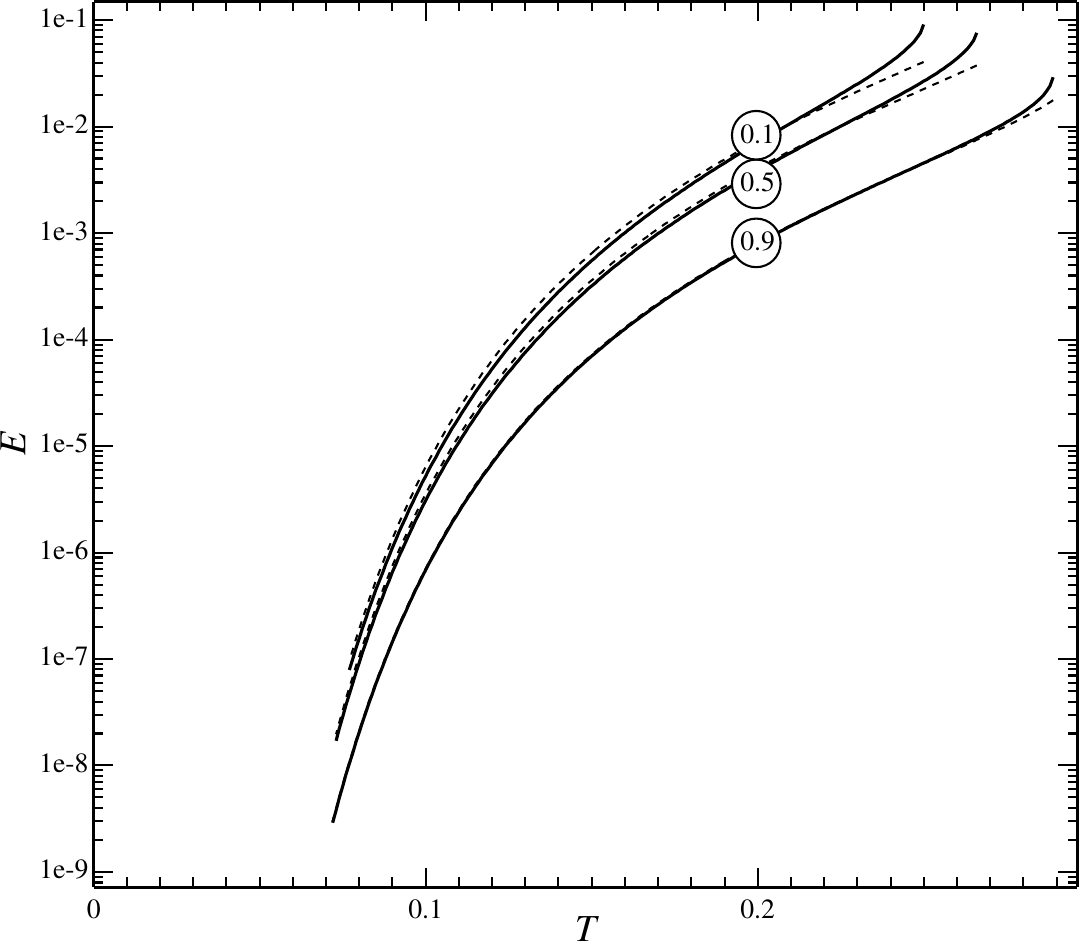}
\caption{Nondimensional evaporation rate $E$ vs. temperature $T$, as described by boundary-value problem (\ref{3.2}) and (\ref{3.4}), (\ref{3.6})--(\ref{3.7}). The curves are marked with the corresponding values of the relative humidity, $\rho^{(v)}/\rho^{(v.sat)}$. The dotted curves show the corresponding asymptotic result, (\ref{3.17})--(\ref{3.19}).}
\label{fig5}
\end{figure}

The dependence of $E$ on $T$ is illustrated in Fig. \ref{fig5}. Note that the
curves depicted could not be extended to $T=0$ due to computational
difficulties (the vapor density becomes too small). In the opposite limit, the
curves are truncated due to the nonexistence of solution of Eq. (\ref{3.8}) as
explained above.

\subsection{The limit of nearly-saturated vapor\label{Sec 3.2}}

If the vapor is close to saturation, evaporation must be slow, i.e.,%
\[
\frac{\rho^{(v)}}{\rho^{(v.sat)}}\approx1\qquad\Rightarrow\qquad E\ll1.
\]
In this case, boundary-value problem (\ref{3.2}), (\ref{3.4}), (\ref{3.6}%
)--(\ref{3.7}) can be solved asymptotically.

If the fluid temperature is far from its critical value, the problem involves
another small parameter, $\rho^{(v.sat)}/\rho^{(l.sat)}$, making the analysis
awkward. To remedy this, it is assumed that this parameter is order-one, then
the asymptotic results are shown numerically to work for $\rho^{(v.sat)}%
/\rho^{(l.sat)}\ll1$ as well.

Define the equilibrium solution $\rho^{(0)}(z)$ as that of Eq. (\ref{3.6})
with $E=0$, subject to boundary conditions (\ref{3.2}), (\ref{3.4}), and
(\ref{3.7}) with $\rho^{(l)}=\rho^{(l.sat)}$ and $\rho^{(v)}=\rho^{(v.sat)}$,
i.e.,%
\begin{multline}
\rho^{(0)}\frac{\mathrm{d}^{2}\rho^{(0)}}{\mathrm{d}\xi^{2}}-\frac{1}%
{2}\left(  \frac{\mathrm{d}\rho^{(0)}}{\mathrm{d}\xi}\right)  ^{2}\\
-p(\rho^{(0)},T)+p(\rho^{(v.sat)},T)=0, \label{3.10}%
\end{multline}%
\begin{equation}
\rho^{(0)}\rightarrow\rho^{(l.sat)}\qquad\text{as}\qquad\xi\rightarrow-\infty,
\label{3.11}%
\end{equation}%
\begin{equation}
\rho^{(0)}\rightarrow\rho^{(v.sat)}\qquad\text{as}\qquad\xi\rightarrow+\infty,
\label{3.12}%
\end{equation}%
\begin{equation}
\rho^{(0)}=\frac{\rho^{(l.sat)}+\rho^{(v.sat)}}{2}\qquad\text{at}\qquad\xi=0.
\label{3.13}%
\end{equation}
It can be shown that, unless $\rho^{(l.sat)}$ and $\rho^{(v.sat)}$ satisfy the
Maxwell construction (\ref{2.15})--(\ref{2.16}), the above boundary-value
problem does not have a solution (see Appendix \ref{Appendix B.1}).

The solution $\rho$ of the full problem (\ref{3.2}), (\ref{3.4}),
(\ref{3.6})--(\ref{3.7}) is close, but not equal, to the equilibrium solution
$\rho^{(0)}$. The deviation of the former from the latter is comparable to the
evaporation rate $E$, so let%
\[
\rho=\rho^{(0)}+\rho^{(1)}+\cdots,
\]
where $\rho^{(1)}=\mathcal{O}(E)$ and $\cdots=\mathcal{O}(E^{2})$. Assume also
that $E$ scales with the deviation of the relative humidity from unity,%
\[
1-\frac{\rho^{(v)}}{\rho^{(v.sat)}}=\mathcal{O}(E),
\]
and that the liquid density is close to its saturated value by the same
margin,%
\[
1-\frac{\rho^{(l)}}{\rho^{(l.sat)}}=\mathcal{O}(E).
\]
To zeroth order, Eq. (\ref{3.10}) and boundary conditions (\ref{3.2}),
(\ref{3.4}) are satisfied identically. The first order yields, after
straightforward algebra,\begin{widetext}%
\begin{equation}
\frac{\mathrm{d}}{\mathrm{d}\xi}\left(  \frac{1}{\rho^{(0)}}\frac
{\mathrm{d}\rho^{(1)}}{\mathrm{d}\xi}\right)  +\frac{1}{\rho^{(0)2}}\left[
\frac{\mathrm{d}^{2}\rho^{(0)}}{\mathrm{d}\xi ^{2}}-\frac{\partial p(\rho
^{(0)},T)}{\partial\rho^{(0)}}\right]  \rho^{(1)}=\frac{E\eta(\rho^{(0)}%
,T)}{\rho^{(0)4}}\frac{\mathrm{d}\rho^{(0)}}{\mathrm{d}\xi}-\frac{1}%
{\rho^{(0)2}}\frac{\partial p(\rho^{(v.sat)},T)}{\partial\rho^{(v.sat)}%
}\left(  \rho^{(v)}-\rho^{(v.sat)}\right)  ,\label{3.14}%
\end{equation}
\end{widetext}%
\begin{equation}
\rho^{(1)}\rightarrow\rho^{(l)}-\rho^{(l.sat)}\qquad\text{as}\qquad
\xi\rightarrow-\infty, \label{3.15}%
\end{equation}%
\begin{equation}
\rho^{(1)}\rightarrow\rho^{(v)}-\rho^{(v.sat)}\qquad\text{as}\qquad
\xi\rightarrow+\infty. \label{3.16}%
\end{equation}
The evaporation rate $E$ can be found without finding $\rho^{(1)}(\xi)$ (and
even considering $\rho^{(2)}(\xi)$ and the higher corrections). To do so,
observe that the operator on the left-hand side of Eq. (\ref{3.14}) is
self-adjoint, and the homogeneous version of problem (\ref{3.14}%
)--(\ref{3.16}) is satisfied by $\rho^{(1)}=\mathrm{d}\rho^{(0)}/\mathrm{d}%
\xi$ -- hence, the full (nonhomogeneous) version should be orthogonal to
$\mathrm{d}\rho^{(0)}/\mathrm{d}\xi$.

To derive the orthogonality condition, multiply (\ref{3.14}) by $\mathrm{d}%
\rho^{(0)}/\mathrm{d}\xi$ and integrate from $\xi=-\infty$ to $\xi=+\infty$.
Using boundary condition (\ref{3.11})--(\ref{3.12}) for $\rho^{(0)}$ and
(\ref{3.15})--(\ref{3.16}) for $\rho^{(1)}$, one obtains the desired
expression for the evaporation rate:%
\begin{equation}
E=\frac{H}{A}, \label{3.17}%
\end{equation}
where%
\begin{align}
H &  =\left(  1-\frac{\rho^{(v.sat)}}{\rho^{(l.sat)}}\right)  \frac{\partial
p(\rho^{(v.sat)},T)}{\partial\rho^{(v.sat)}}\frac{\rho^{(v.sat)}-\rho^{(v)}%
}{\rho^{(v.sat)}},\label{3.18}\\
A &  =\int_{-\infty}^{\infty}\frac{\eta(\rho^{(0)},T)}{\rho^{(0)4}}\left(
\frac{\mathrm{d}\rho^{(0)}}{\mathrm{d}\xi}\right)  ^{2}\mathrm{d}%
\xi.\label{3.19}%
\end{align}
Observe that the last factor in the expression for $H$ is the difference
between the relative humidity and unity.

Asymptotic result (\ref{3.17})--(\ref{3.19}) is compared to the numeric
solution of the exact problem in Fig. \ref{fig5}. Interestingly, the agreement
between the two solutions is reasonably good even when the relative humidity
is far from unity: the relative error of the most part of the asymptotic curve
for the 50\% humidity is less than $0.15$ and that for the 10\% curve is less
than $0.22$. These errors are only exceeded near the terminal point, where the
asymptotic solution noticeably under-predicts the exact one.

The coefficient $A$ given by expression (\ref{3.19}) always arises when the
DIM is used to examine evaporation \cite{Benilov22a}, condensation
\cite{Benilov22b}, or another setting where these processes play a role
\cite{Benilov20d}. It involves the leading-order solution $\rho^{(0)}(z)$
which needs to be computed before the integral in (\ref{3.19}) can be
evaluated. To avoid this extra computation, one can express $A$ as a
closed-form integral (see Appendix \ref{Appendix B.2})\begin{widetext}%
\begin{equation}
A=2^{1/2}\int_{\rho^{(v.sat)}}^{\rho^{(l.sat)}}\frac{\eta(\rho,T)}{\rho^{4}%
}\sqrt{\rho\left[  G(\rho,T)-G(\rho^{(v.sat)},T)\right]  -p(\rho
,T)+p(\rho^{(v.sat)},T)}\mathrm{d}\rho.\label{3.20}%
\end{equation}
\end{widetext}If the nondimensional temperature is low (which is the case for
many common fluids at \textquotedblleft normal conditions\textquotedblright%
\ \cite{Benilov20b}), expressions (\ref{3.20}) can be calculated asymptotically,%

\begin{equation}
A\approx0.14219\frac{\eta(0,T)\,T^{1/2}}{\left(  \rho^{(v.sat)}\right)
^{5/2}}\qquad\text{if}\qquad T\ll1, \label{3.21}%
\end{equation}
where $\eta(0,T)$ is the small-density limit of the viscosity $\eta(\rho,T)$.
Since $\rho^{(v.sat)}\rightarrow0$ as $T\rightarrow0$, one can deduce from the
above expression and formula (\ref{3.17}) that the evaporation rate vanishes
as $T\rightarrow0$ (as it should do physically).

\subsection{Unsteady evaporation\label{Sec 3.3}}

It remains to find out how the liquid evaporates if no steady solution exists
. In terms of Fig. \ref{fig3}b, this occurs when the pair $(T,\rho^{(v)})$ is
outside the existence region.

\begin{figure}
\includegraphics[width=\columnwidth]{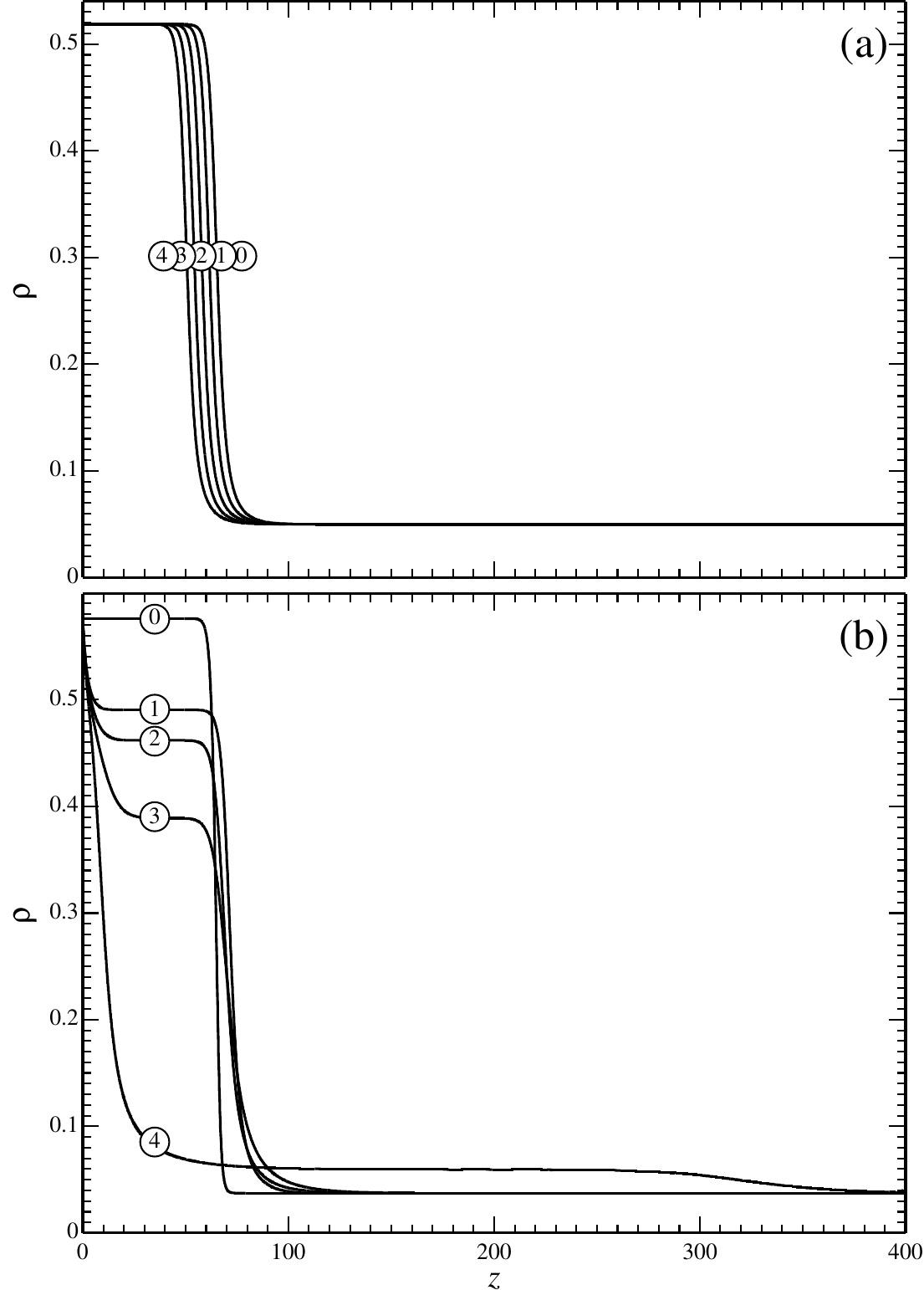}
\caption{Two scenarios of isothermal evaporation. In both cases, $T=0.26$ and the near-substrate density $\rho_{0}$ equals, for simplicity, the value prescribed by the initial condition. (a)~$\rho^{(v)}/\rho^{(v.sat)}=0.4$ (a steady solution exists and is used as the initial condition); (b)~$\rho^{(v)}/\rho^{(v.sat)}=0.3$ (steady solution does not exist, the initial condition is given by (\ref{3.21})--(\ref{3.22})]. In both cases, the time $t$ of the snapshot is related to the curve number $n$ by $t=22\,n$.}
\label{fig6}
\end{figure}

The steady and unsteady scenarios of evaporation were simulated using the set
of evolution equations (\ref{2.27})--(\ref{2.28}) using the method of lines
\cite{Schiesser78}. Typical results are illustrated in Fig. \ref{fig6}.

The parameters of Fig. \ref{fig6}(a) are such that a steady solution exists.
Even though it was obtained for an unbounded space, the presence of the
substrate does not change it until the interface is very close to the
substrate (this stage of the evolution is not shown in the figure). The run
depicted in Fig. \ref{fig6}(a) originates from an initial condition obtained
by solving the steady problem (\ref{3.2}), (\ref{3.4}), (\ref{3.6}%
)--(\ref{3.7}) and shifting the solution to the right by a distance of $\Delta
z=65$ from the substrate.

For the parameters of Fig. \ref{fig6}(b), no steady solution exists. The
following initial condition was used:%
\begin{multline}
\rho=\rho^{(0)}(z-\Delta z)\left[  \left(  \frac{\rho^{(v)}}{\rho^{(v.sat)}%
}+1\right)  \right. \\
+\left.  \left(  \frac{\rho^{(v)}}{\rho^{(v.sat)}}-1\right)  \tanh
\frac{z-z_{1}}{W}\right]  , \label{3.22}%
\end{multline}
where $\rho^{(0)}(z)$ is the equilibrium solution described by (\ref{3.10}%
)--(\ref{3.13}), $\Delta z$ is the initial position of the interface, $z_{1}$
is the position of a transitional region where the vapor density changes from
$\rho^{(v.sat)}$ [as `prescribed' by $\rho^{(0)}(z)$] to $\rho^{(v)}$
corresponding to the chosen relative humidity, and $W$ is this region's width
. In Fig. \ref{fig6}(b), the following parameter values were used:%
\begin{equation}
\Delta z=65,\qquad z_{1}=67,\qquad W=2. \label{3.23}%
\end{equation}
The difference between the two scenarios is evident: if no steady solution
exists, the whole layer `empties out' -- quickly and all at once. This occurs
because the pressure in the liquid cannot equilibrate with that in the vapor
(recall that condition (\ref{3.8}) does not hold) -- and the resulting
pressure gradient generates a strong evaporative flow. It can be conjectured
that, in a three-dimensional model -- where bubbles may arise -- the liquid in
this regime \emph{boils}. If this conjecture is true, the impossibility of
matching the pressure in the vapor and liquid phases at a certain temperature
provides a criterion for a `near-vacuum boiling'\ (which includes the `vacuum
boiling' proper\ as a limiting case).

Observe the near-substrate boundary layer in Fig. \ref{fig6}(b). It develops
due to boundary condition (\ref{2.22}) forcing the near-substrate density to
remain fixed. Computations with various values of $\rho_{0}$ have shown that
this boundary layer has no impact on the global dynamics.

Note that evaporation of a pure fluid has been previously examined in Ref.
\cite{BarbanteFrezzottiGibelli14} -- and not only via the DIM, but also via
simulations molecular dynamics. The parameter range explored in this paper
happened to be inside the existence region of steady solutions, so the
unsteady regime was not observed.

\section{Nonisothermal evaporation\label{Sec 4}}

If evaporation is steady, the density field near the interface is microscopic
-- i.e., its dimensional spatial scale is comparable to $l$ introduced in Sec.
\ref{Sec 2.3}. For the temperature field, however, no mechanism exists for
keeping it short-scale. As a result, evaporative cooling rapidly spreads out
and eventually become macroscopic.

Thus, the problem splits into two subproblems:\smallskip

(A) an analysis of the \emph{macroscopic} temperature field, which yields the
heat fluxes coming into the interface from the vapor and liquid
sides,\smallskip

(B) an analysis of the \emph{microscopic} interfacial dynamics, which
inter-relates the heat fluxes and evaporation rate.\smallskip

\noindent Subproblems (A) and (B) are examined in Secs. \ref{Sec 4.1} and
\ref{Sec 4.2}, respectively. The former is solved three times: for an
unbounded space (the simplest case), for a semi-infinite space bounded by a
fixed-temperature substrate (the case with most applications), and for an
insulated substrate.

Admittedly, the analyses presented below involve several assumptions, so the
results obtained will be verified via numerical simulations of the exact
governing equations in Sec. \ref{Sec 4.3}.

\subsection{Macroscopic solution\label{Sec 4.1}}

The evaporative cooling is unlikely to exceed, say, 10--20 degrees --
otherwise, as argued in the Introduction, the evaporation would effectively
stop. Within the normal-conditions\ range, 20 degrees is a small fraction of
the absolute temperature, so that $T(z,t)$ can be decomposed into a constant
part $T_{0}$ and a small variation $\tilde{T}(z,t)$,%
\begin{equation}
T=T_{0}+\tilde{T}.\label{4.1}%
\end{equation}
For a fixed-$T$ substrate, $T_{0}$ is its temperature.

Consider liquid and vapor far from the interface, were the density is close to
either $\rho^{(l)}$ or $\rho^{(v)}$. The spatial scale of the temperature
field in these regions is macroscopic, which nondimensionally means
\textquotedblleft much greater than unity\textquotedblright. Under such
assumptions, the governing equations (\ref{2.27})--(\ref{2.29}) can be reduced
to a heat conduction equation (see Appendix \ref{C.2}).

Let the position of the interface be $z=z_{i}(t)$. Its velocity can be
expressed in terms of the evaporation rate and the liquid's density,%
\[
\frac{\mathrm{d}z_{i}}{\mathrm{d}t}=-\frac{E}{\rho^{(l)}}.
\]
Changing to the co-moving reference frame $\left(  z,t\right)  \rightarrow
\left(  \xi,t\right)  $, where%
\begin{equation}
\xi=z-z_{i}=z+\int\frac{E(t)}{\rho^{(l)}}\mathrm{d}t,\label{4.2}%
\end{equation}
one can describe the temperature evolution by%
\begin{align}
\rho^{(l)}c_{P}^{(l)}\left(  \frac{\partial\tilde{T}}{\partial t}+\frac
{E}{\rho^{(l)}}\frac{\partial\tilde{T}}{\partial z}\right)   &  =\kappa
^{(l)}\frac{\partial^{2}\tilde{T}}{\partial z^{2}}\hspace{0.4cm}%
\,\text{if}\hspace{0.4cm}z<0,\label{4.3}\\
\rho^{(v)}c_{P}^{(v)}\left(  \frac{\partial\tilde{T}}{\partial t}+\frac
{E}{\rho^{(l)}}\frac{\partial\tilde{T}}{\partial z}\right)   &  =\kappa
^{(v)}\frac{\partial^{2}\tilde{T}}{\partial z^{2}}\hspace{0.4cm}%
\text{if}\hspace{0.4cm}z>0,\label{4.4}%
\end{align}
where%
\[
c_{P}^{(l)}=c_{P}(\rho^{(l)},T_{0}),\qquad c_{P}^{(v)}=c_{P}(\rho^{(v)}%
,T_{0}),
\]%
\[
\kappa^{(l)}=\kappa(\rho^{(l)},T_{0}),\qquad\kappa^{(v)}=\kappa(\rho
^{(v)},T_{0}),
\]
Eqs. (\ref{4.3})--(\ref{4.4}) are to be solved with the following conditions:%
\begin{equation}
\left(  \tilde{T}\right)  _{z=+0}=\left(  \tilde{T}\right)  _{z=-0}%
,\label{4.5}%
\end{equation}%
\begin{equation}
\kappa^{(v)}\left(  \frac{\partial\tilde{T}}{\partial z}\right)
_{z=+0}-\kappa^{(l)}\left(  \frac{\partial\tilde{T}}{\partial z}\right)
_{z=-0}=E\,\Delta h,\label{4.6}%
\end{equation}
where the right-hand side of (\ref{4.6}) describes consumption of heat in the
interface due to evaporative cooling. The simplest initial condition will be
assumed,%
\begin{equation}
\tilde{T}=0\qquad\text{at}\qquad t=0,\label{4.7}%
\end{equation}
i.e., initially, the temperature field is uniform.

The problem formulated contains a hidden small parameter. To identify it,
denote the vertical spatial scale of the solution by $D$ (in case of a finite
layer, $D$ coincides with its depth) and observe that%
\begin{align}
\text{Eq. (\ref{4.3})}  &  \text{:}\qquad\frac{E\text{-involving term}%
}{\text{right-hand side}}\sim ED\frac{c_{P}^{(l)}}{\kappa^{(l)}},\label{4.8}\\
\text{Eq. (\ref{4.4})}  &  \text{:}\qquad\frac{E\text{-involving term}%
}{\text{right-hand side}}\sim ED\frac{c_{P}^{(v)}}{\kappa^{(v)}}\frac
{\rho^{(v)}}{\rho^{(l)}}. \label{4.9}%
\end{align}
Thus, if%
\begin{equation}
D\ll\frac{\kappa^{(l)}}{Ec_{P}^{(l)}}, \label{4.10}%
\end{equation}
the $E$-involving term in Eq. (\ref{4.3}) can be neglected.

To understand how restrictive condition (\ref{4.10}) is, one can estimate its
right-hand side for water at, say, $25^{\circ}\mathrm{C}$. Using Ref.
\cite{LindstromMallard97} to obtain values for $\kappa^{(l)}$ and $c_{P}%
^{(l)}$ and assuming estimate (\ref{1.4}) for $E$, one obtains%
\[
D\ll5.7\,\mathrm{m}.
\]
Many physically important examples comply with this restriction, and those
that do not are discussed in Sec. \ref{Sec 4.3}.

Observe also the small factor $\rho^{(v)}/\rho^{(l)}$ in estimate (\ref{4.9})
-- hence, its right-hand is likely to be smaller than that of (\ref{4.8}).
Estimating the ratio of the two right-hand sides for water and air at
$25^{\circ}\mathrm{C}$, one obtains%
\[
\frac{\left(  c_{P}^{(v)}/\kappa^{(v)}\right)  \left(  \rho^{(v)}/\rho
^{(l)}\right)  }{\left(  c_{P}^{(l)}/\kappa^{(l)}\right)  }\approx0.0066.
\]
Thus, if the $E$-involving term in\ Eq. (\ref{4.3}) is negligible, its
counterpart in Eq. (\ref{4.4}) is too.

Omitting these terms in Eqs. (\ref{4.3})--(\ref{4.4}), one obtains%
\begin{align}
\rho^{(l)}c_{P}^{(l)}\frac{\partial\tilde{T}}{\partial t}  &  =\kappa
^{(l)}\frac{\partial^{2}\tilde{T}}{\partial z^{2}}\qquad\,\text{if}\qquad
z<0,\label{4.11}\\
\rho^{(v)}c_{P}^{(v)}\frac{\partial\tilde{T}}{\partial t}  &  =\kappa
^{(v)}\frac{\partial^{2}\tilde{T}}{\partial z^{2}}\qquad\text{if}\qquad z>0.
\label{4.12}%
\end{align}
These equations do not explicitly involve $t$ -- hence, can be solved via the
Laplace transformation (the presence of $E(t)$ in the matching condition
(\ref{4.6}) does not pose a problem, as this is not a coefficient but a
right-hand side).

\subsubsection{A flat interface in an unbounded space\label{Sec 4.1.1}}

This case corresponds to the following boundary conditions:%
\begin{equation}
\tilde{T}\rightarrow0\qquad\text{as}\qquad z\rightarrow\pm\infty.\label{4.13}%
\end{equation}
Applying the Laplace transformation to Eqs. (\ref{4.11})--(\ref{4.12}), then
recalling initial condition (\ref{4.7}) and boundary conditions (\ref{4.5}%
)--(\ref{4.6}) and (\ref{4.13}), one obtains\begin{widetext}%
\begin{align*}
\hat{T} &  =-\frac{s^{-1/2}\hat{E}\,\Delta h}{\sqrt{\rho^{(l)}c_{P}%
^{(l)}\kappa^{(l)}}+\sqrt{\rho^{(v)}c_{P}^{(v)}\kappa^{(v)}}}\exp\left(
\sqrt{\frac{\rho^{(l)}c_{P}^{(l)}s}{\kappa^{(l)}}}z\right)  \hspace
{1.15cm}\text{if}\qquad z<0,\\
\hat{T} &  =-\frac{s^{-1/2}\hat{E}\,\Delta h}{\sqrt{\rho^{(l)}c_{P}%
^{(l)}\kappa^{(l)}}+\sqrt{\rho^{(v)}c_{P}^{(v)}\kappa^{(v)}}}\exp\left(
-\sqrt{\frac{\rho^{(v)}c_{P}^{(v)}s}{\kappa^{(v)}}}z\right)  \qquad
\text{if}\qquad z>0,
\end{align*}
\end{widetext}where the variables with hats represent the transforms, e.g.,%
\[
\hat{T}(s,t)=\int_{0}^{\infty}\tilde{T}(z,t)\operatorname{e}^{-st}\mathrm{d}t.
\]
The inverse transform of $\hat{T}(s,t)$ will be matched to the microscopic
solution obtained later. Three characteristics of the former will be needed
for the matching: the temperature at the interface and the heat fluxes toward
it. Calculating the inverse transforms of these characteristics only [which is
easier than calculating the whole $T(z,t)$], one obtains\begin{widetext}%
\begin{equation}
\left(  \tilde{T}\right)  _{z=0}=-\frac{\Delta h}{\sqrt{\rho^{(l)}c_{P}%
^{(l)}\kappa^{(l)}}+\sqrt{\rho^{(v)}c_{P}^{(v)}\kappa^{(v)}}}\int_{0}^{t}%
\frac{E(t^{\prime})}{\sqrt{\pi\left(  t-t^{\prime}\right)  }}\mathrm{d}%
t^{\prime},\label{4.14}%
\end{equation}%
\begin{equation}
\kappa^{(l)}\left(  \frac{\partial\tilde{T}}{\partial z}\right)
_{z=-0}=-\frac{\sqrt{\rho^{(l)}c_{P}^{(l)}\kappa^{(l)}}E\,\Delta h}{\sqrt
{\rho^{(l)}c_{P}^{(l)}\kappa^{(l)}}+\sqrt{\rho^{(v)}c_{P}^{(v)}\kappa^{(v)}}%
},\qquad\kappa^{(v)}\left(  \frac{\partial\tilde{T}}{\partial z}\right)
_{z=+0}=\frac{\sqrt{\rho^{(v)}c_{P}^{(v)}\kappa^{(v)}}E\,\Delta h}{\sqrt
{\rho^{(l)}c_{P}^{(l)}\kappa^{(l)}}+\sqrt{\rho^{(v)}c_{P}^{(v)}\kappa^{(v)}}%
}.\label{4.15}%
\end{equation}
\end{widetext}Note that (\ref{4.15}) [but not (\ref{4.14})] can be deduced
from basic symmetries of the heat-conduction equations (\ref{4.11}%
)--(\ref{4.12}) -- which probably explains their relatively simple form.

\subsubsection{Fixed-temperature substrate\label{Sec 4.1.2}}

In the coordinate system co-moving with the interface, a fixed-$T$ substrate
corresponds to the following boundary condition,%
\begin{equation}
\tilde{T}=0\qquad\text{at}\qquad z=-D,\label{4.16}%
\end{equation}
where $D(t)$ is the depth of the liquid layer. Since this condition is set at
a $\emph{moving}$ boundary, Eqs. (\ref{4.11})--(\ref{4.12}) can no longer be
solved exactly, but one can still solve them asymptotically under the same
assumption (\ref{4.10}) which makes the $E$-involving term negligible. It
makes evaporation slower than the temperature evolution -- hence, one can
`freeze' $D$ -- i.e., assume that it depends on a slow-time variable different
from $t$.

Now, apply the Laplace transformation to Eqs. (\ref{4.11})--(\ref{4.12}) and
boundary conditions (\ref{4.5})--(\ref{4.6}), (\ref{4.16}) with `frozen' $D$.
After straightforward algebra, one can show that the temperature field tends
to\begin{widetext}%
\begin{equation}
\tilde{T}\rightarrow\left\{
\begin{tabular}
[c]{ll}%
$-\dfrac{E\,\Delta h}{\kappa^{(l)}}\left(  z+D\right)  \qquad$ &
$\text{if}\qquad-D<z<0,$\\
$-\dfrac{E\,\Delta h}{\kappa^{(l)}}D$ & $\text{if}\qquad z>0$%
\end{tabular}
\right.  \qquad\text{as}\qquad t\rightarrow\infty.\label{4.17}%
\end{equation}
\end{widetext}Solution (\ref{4.17}) has a clear physical meaning: it describes
a quasi-steady heat flux from the substrate toward the interface, feeding
evaporation. The vapor above the interface has uniform temperature because the
temperature variations have spread out to $z\rightarrow+\infty$.

Note that the linear dependence of $T$ on $z$ in liquid has indeed been
observed experimentally \cite{WardStanga01}.

\subsubsection{Thermally insulated substrate\label{Sec 4.1.3}}

The boundary condition at the substrate in this case is%
\[
\frac{\partial\tilde{T}}{\partial z}=0\qquad\text{at}\qquad z=-D.
\]
As before, the initial-boundary-value problem for $\tilde{T}(z,t)$ (with
$D(t)$ `frozen'), can be solved via the Laplace transformation -- but the
resulting solution is much more cumbersome than those in the two previous
cases. One can still see that, in the liquid, the temperature field becomes
spatially uniform, with a small parabolic correction:%
\begin{multline}
\tilde{T}\approx\tilde{T}_{0}(t)\\
+\frac{\mathrm{d}\tilde{T}_{0}}{\mathrm{d}t}\frac{\rho^{(l)}c_{P}^{(l)}%
}{\kappa^{(l)}}\frac{\left(  D+z\right)  ^{2}}{2}\qquad\text{if}\qquad-D<z<0.
\label{4.18}%
\end{multline}
Since the case of insulated substrate is cumbersome mathematically and trivial
physically, it will not be examined analytically in further detail (but will
be simulated numerically in Sec. \ref{Sec 4.3}).

\subsection{Microscopic solution\label{Sec 4.2}}

Strictly speaking, nonisothermal evaporation is not steady, but can be
regarded as \emph{quasi}steady -- i.e., adjusted to the current temperature of
the interface and incoming heat fluxes, and slowly changing with them.

In this section, $E$ is calculated asymptotically for quasi-steady
nonisothermal evaporation under an additional assumption that the relative
humidity is close to 100\%. Only the case of fixed-$T$ substrate will be
examined; those of insulated substrate and unbounded space will be briefly
discussed in the next subsection.

The assumption of quasi-steadiness amounts to rewriting the governing
equations (\ref{2.27})--(\ref{2.29}) in the co-moving reference frame $\left(
\xi,t\right)  $ [$\xi$ is defined by (\ref{4.2})] and omitting the time
derivatives, which yields%
\begin{equation}
\frac{E}{\rho^{(l)}}\frac{\partial\rho}{\partial\xi}+\frac{\partial\left(
\rho w\right)  }{\partial\xi}=0,\label{4.19}%
\end{equation}%
\begin{equation}
p=\eta\frac{\partial w}{\partial\xi}+\rho\frac{\partial^{2}\rho}{\partial
\xi^{2}}-\frac{1}{2}\left(  \frac{\partial\rho}{\partial\xi}\right)
^{2}+p^{(v)},\label{4.20}%
\end{equation}%
\begin{multline}
\frac{E}{\rho^{(l)}}\frac{\partial\left(  \rho e\right)  }{\partial\xi}%
+\frac{\partial}{\partial\xi}\left[  w\left(  \rho e+p-\eta\frac{\partial
w}{\partial\xi}\right)  -\kappa\frac{\partial T}{\partial\xi}\right]  \\
=w\rho\frac{\partial^{3}\rho}{\partial\xi^{3}}.\label{4.21}%
\end{multline}
These equations are to be solved with the `old' boundary conditions for $w$,
(\ref{3.1}) and (\ref{3.3}), whereas those for $\rho$ and $T$ will be obtained
later via matching with the macroscopic solution.

As in the isothermal case, one can reduce Eq. (\ref{4.19}) to expression
(\ref{3.5}) for $w$, and substitute it into Eqs. (\ref{4.20})--(\ref{4.21})
which become\begin{widetext}%
\begin{equation}
p(\rho,T)=-\frac{E\eta(\rho,T)}{\rho^{2}}\frac{\partial\rho}{\partial\xi}%
+\rho\frac{\partial^{2}\rho}{\partial\xi^{2}}-\frac{1}{2}\left(
\frac{\partial\rho}{\partial\xi}\right)  ^{2}+p(\rho^{(v)},T_{0}),\label{4.22}%
\end{equation}%
\begin{equation}
\kappa(\rho,T)\frac{\partial T}{\partial\xi}=E\left[  e(\rho,T)+\left(
1-\frac{\rho}{\rho^{(l)}}\right)  \frac{p(\rho,T)}{\rho}-\left(  1-\frac{\rho
}{\rho^{(l)}}\right)  \frac{\partial^{2}\rho}{\partial\xi^{2}}-\frac{1}%
{2\rho^{(l)}}\left(  \frac{\partial\rho}{\partial\xi}\right)  ^{2}-Q\right]
+E^{2}\left(  \frac{1}{\rho}-\frac{1}{\rho^{(l)}}\right)  \frac{\eta}{\rho
^{2}}\frac{\partial\rho}{\partial\xi},\label{4.23}%
\end{equation}
where Eq. (\ref{4.23}) was integrated and $Q$ is the integration constant
(to be fixed later).
As before, assume that the solution is close to equilibrium: the leading-order
temperature is uniform and equals $T_{0}$ [the same constant as in the
macroscopic representation (\ref{4.1})], and the leading-order density is
described by the solution $\rho^{(0)}$ of the equilibrium boundary-value
problem (\ref{3.10})--(\ref{3.13}). These assumptions amount to%
\[
\rho=\rho^{(0)}+\rho^{(1)}+\mathcal{\cdots},\qquad T=T_{0}+T^{(1)}+\cdots,
\]
where $\rho^{(1)}$ and $T^{(1)}$ are $\mathcal{O}(E)$.
Upon substitution of these expansions into Eq. (\ref{4.22}), the zeroth order
cancels out and the first-order yields%
\begin{multline}
\frac{\partial}{\partial\xi}\left(  \frac{1}{\rho^{(0)}}\frac{\partial^{2}%
\rho^{(1)}}{\partial\xi^{2}}\right)  +\frac{1}{\rho^{(0)2}}\left[
\frac{\partial^{2}\rho^{(0)}}{\partial\xi^{2}}-\frac{\partial p(\rho
^{(0)},T_{0})}{\partial\rho^{(0)}}\right]  \rho^{(1)}\\
=\frac{E\eta(\rho^{(0)},T_{0})}{\rho^{4}}\frac{\partial\rho^{(0)}}{\partial
\xi}-\frac{1}{\rho^{(0)2}}\frac{\partial p(\rho^{(v.sat)},T_{0})}{\partial
\rho^{(v.sat)}}\left(  \rho^{(v)}-\rho^{(v.sat)}\right)  +\frac{1}{\rho
^{(0)2}}\frac{\partial p(\rho^{(0)},T_{0})}{\partial T_{0}}T^{(1)}.
\label{4.24}%
\end{multline}
By comparison with its isothermal counterpart (\ref{3.14}), Eq. (\ref{4.24})
includes an extra term (the last term on its right-hand side).
Expanding, in turn, the temperature equation (\ref{4.23}), one obtains%
\[
\kappa(\rho^{(0)},T_{0})\frac{\partial T^{(1)}}{\partial\xi}=E\left[
e(\rho^{(0)},T_{0})+\left(  \frac{1}{\rho^{(0)}}-\frac{1}{\rho^{(l.sat)}%
}\right)  p(\rho^{(0)},T_{0})-\left(  1-\frac{\rho^{(0)}}{\rho^{(l.sat)}%
}\right)  \frac{\partial^{2}\rho^{(0)}}{\partial\xi^{2}}-\frac{1}%
{2\rho^{(l.sat)}}\left(  \frac{\partial\rho^{(0)}}{\partial\xi}\right)
^{2}-Q\right]  .
\]
The derivatives of $\rho^{(0)}$ in this equation can be eliminated via
equalities (\ref{B.1}) and (\ref{B.3}), and after straightforward algebra
involving the use of definition (\ref{2.3}) of the chemical potential, one
obtains%
\begin{equation}
\kappa(\rho^{(0)},T_{0})\frac{\partial T^{(1)}}{\partial\xi}=ET_{0}\left[
s(\rho^{(0)},T_{0})-s(\rho^{(v.sat)},T_{0})\right]  +e(\rho^{(v.sat)}%
,T_{0})-Q.\label{4.25}%
\end{equation}
\end{widetext}The solution of this equation and the outer (macroscopic)
solution (\ref{4.17}) match only if the former satisfy the following boundary
condition%
\begin{align}
\frac{\partial T^{(1)}}{\partial\xi} &  \rightarrow-\frac{E\,\Delta h(T_{0}%
)}{\kappa(\rho^{(l.sat)},T_{0})}\hspace{1.05cm}\text{as}\qquad\xi
\rightarrow-\infty,\label{4.26}\\
T^{(1)} &  \rightarrow-\dfrac{E\,\Delta h(T_{0})}{\kappa(\rho^{(l.sat)}%
,T_{0})}D\qquad\text{as}\qquad\xi\rightarrow+\infty.\label{4.27}%
\end{align}
Condition (\ref{4.27}) implies that the constant $Q$ in Eq. (\ref{4.25}) is%
\begin{equation}
Q=e(\rho^{(v.sat)},T_{0}),\label{4.28}%
\end{equation}
and condition (\ref{4.26}) yields the same result, but subject to%
\begin{equation}
\Delta h(T_{0})=T_{0}\left[  s(\rho^{(v.sat)},T_{0})-s(\rho^{(l.sat)}%
,T_{0})\right]  ,\label{4.29}%
\end{equation}
which is actually an identity (as shown in Appendix \ref{Appendix D}).

Keeping in mind that asymptotics (\ref{4.26}) predicts linear behavior of
$T^{(1)}(\xi)$ as $\xi\rightarrow-\infty$, one can show that Eq. (\ref{4.24})
admits a similar linear asymptotics for $\rho^{(1)}$, i.e.,%
\begin{equation}
\rho^{(1)}=\mathcal{O}(\xi)\qquad\text{as}\qquad\xi\rightarrow-\infty
.\label{4.30}%
\end{equation}
In vapor, one should, as before, assume%
\begin{equation}
\rho^{(1)}\rightarrow\rho^{(v)}-\rho^{(v.sat)}\qquad\text{as}\qquad
\xi\rightarrow+\infty.\label{4.31}%
\end{equation}
The evaporation rate $E$ can be found using the same procedure as that in the
isothermal case. Multiplying Eq. (\ref{4.24}) by $\mathrm{d}\rho
^{(0)}/\mathrm{d}\xi$, one eliminates $\rho^{(1)}$ by integrating by parts,
and recalling boundary conditions (\ref{4.30})--(\ref{4.31}) for $\rho^{(1)}$
and boundary-value problem (\ref{3.11})--(\ref{3.12}) for $\rho^{(0)}$. After
straightforward algebra involving the use of thermodynamic identity
(\ref{2.5}), one obtains%
\begin{equation}
E=\frac{H}{A+A^{\prime}},\label{4.32}%
\end{equation}
where $H$ and $A$ are given by the isothermal expressions (\ref{3.18}%
)--(\ref{3.19}), respectively, and%
\[
A^{\prime}=-\frac{1}{E}\int_{-\infty}^{\infty}T^{(1)}\frac{\mathrm{d}%
s(\rho^{(0)},T_{0})}{\mathrm{d}\xi}\mathrm{d}\xi
\]
can be shown to be independent of $E$.

To do so, recall that the function $T^{(1)}$ in $A^{\prime}$ is the solution
of boundary-value problem (\ref{4.25})--(\ref{4.28}) which needs to be solved
before $A^{\prime}$ can be calculated. This can be by-passed, however, by
rearranging $A^{\prime}$ in the form\begin{widetext}%
\begin{multline*}
A^{\prime}=-\frac{1}{E}\int_{-\infty}^{\infty}\left[  T^{(1)}+\dfrac{E\,\Delta
h}{\kappa(\rho^{(l.sat)},T_{0})}D\right]  \frac{\mathrm{d}\left[  s(\rho
^{(0)},T_{0})-s(\rho^{(l.sat)},T_{0})\right]  }{\mathrm{d}\xi}\mathrm{d}\xi\\
+\dfrac{\Delta h(T_{0})}{\kappa(\rho^{(l.sat)},T_{0})}D\left[  s(\rho
^{(v.sat)},T_{0})-s(\rho^{(l.sat)},T_{0})\right]  ,
\end{multline*}
then integrating the first term by parts, recalling boundary conditions
(\ref{4.26})--(\ref{4.27}), and rearranging the second term using identity
(\ref{4.29}). As a result, the expression for $A^{\prime}$
involves only $\partial T^{(1)}/\partial\xi$, which can be eliminated using
Eqs. (\ref{4.25}) and (\ref{4.28}). Eventually, one obtains%
\begin{equation}
A^{\prime}=A_{1}^{\prime}+A_{2}^{\prime}D,\label{4.33}%
\end{equation}
where%
\begin{align*}
A_{1}^{\prime}  & =T_{0}\int_{-\infty}^{\infty}\frac{\left[  s(\rho
^{(0)},T_{0})-s(\rho^{(l.sat)},T_{0})\right]  \left[  s(\rho^{(0)}%
,T_{0})-s(\rho^{(v.sat)},T_{0})\right]  }{\kappa(\rho^{(0)},T_{0})}%
\mathrm{d}\xi,\\
A_{2}^{\prime}  & =\dfrac{T_{0}\left[  s(\rho^{(v.sat)},T_{0})-s(\rho
^{(l.sat)},T_{0})\right]  ^{2}}{\kappa(\rho^{(l.sat)},T_{0})}.
\end{align*}
\end{widetext}Expressions (\ref{4.32})--(\ref{4.33}) constitute the most
general asymptotic result of the present work. It can be used to find the
dependence of the evaporation rate $E$ and the liquid's depth $D$ on the time
variable $t$; to do so, one needs to complement (\ref{4.32})--(\ref{4.33})
with the ordinary differential equation%
\[
\frac{\mathrm{d}D}{\mathrm{d}t}=-\frac{E}{\rho^{(l.sat)}}.
\]
With $E$ determined by expressions (\ref{4.32}), the above equation can be
readily solved.

\subsection{Numerical simulations\label{Sec 4.3}}

The above predictions regarding the substrate's effect on evaporation have
been tested via numerical integration of the exact governing equations
(\ref{2.27})--(\ref{2.29}), for various initial conditions and parameter
values. Note that, in the numerics, the energy equation (\ref{2.29}) was
conveniently replaced with the (mathematically equivalent) temperature
equation (\ref{C.1}).

\begin{figure*}
\includegraphics[width=\textwidth]{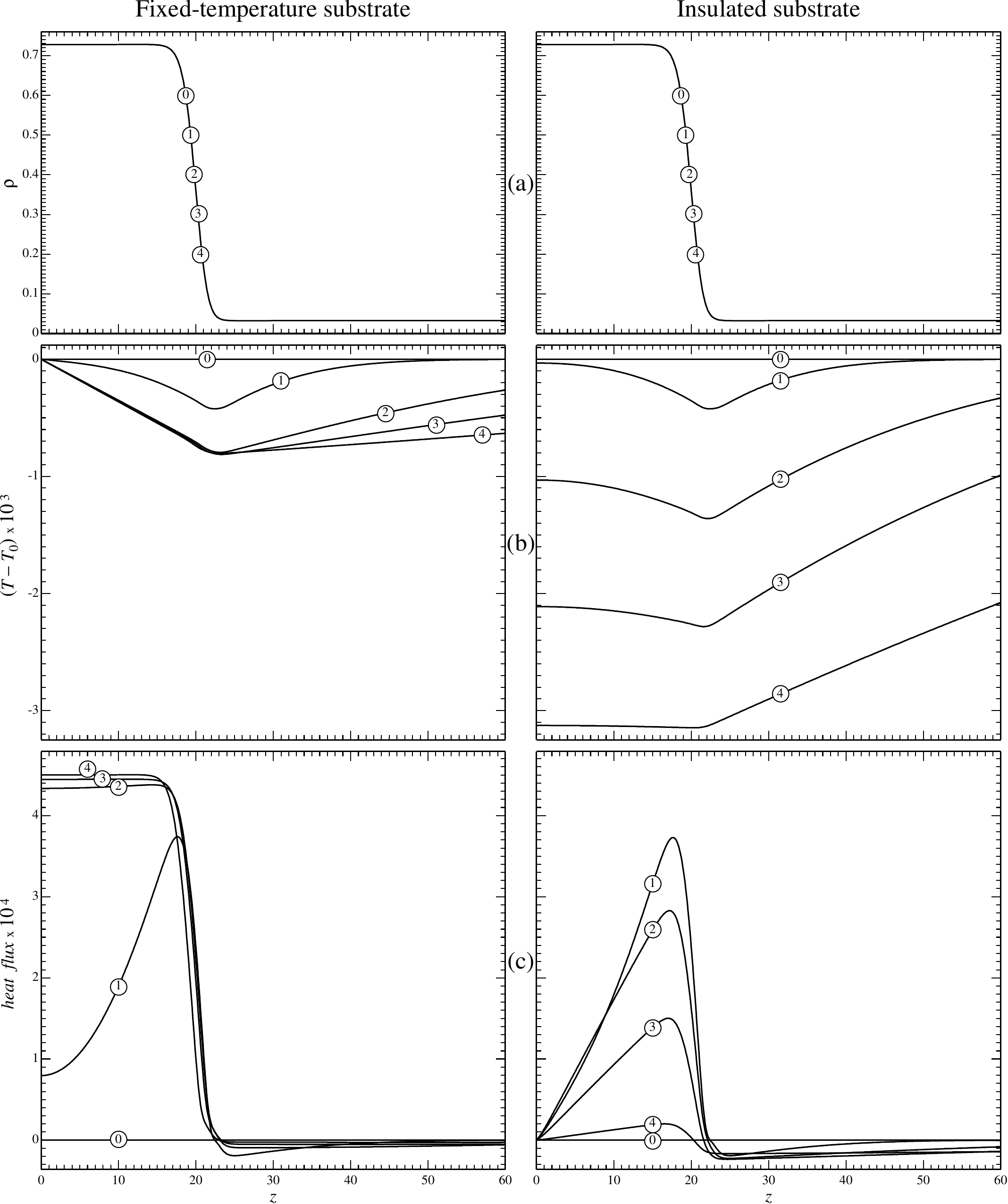}
\caption{Numerical solutions for a fixed-temperature substrate (left-hand panels) and insulated substrate (right-hand panels): (a) density field, (b) temperature field, (c) heat flux. The initial conditions and parameters are described in Sec. \ref{Sec 4.3}. The curves represent the snapshots taken at the following times: (0) $t=0$, (1) $t=20$, (2) $t=200$, (3) $t=500$, (4) $t=1400$. The density distribution evolves very slowly, to the extent that curves (0)--(4) in the upper panels are indistinguishable. Note that this figure shows only a fraction of the $z$ range of the underlying computation which was carried for $z\in\left(  0,400\right)  $.}
\label{fig7}
\end{figure*}

\begin{figure}
\includegraphics[width=\columnwidth]{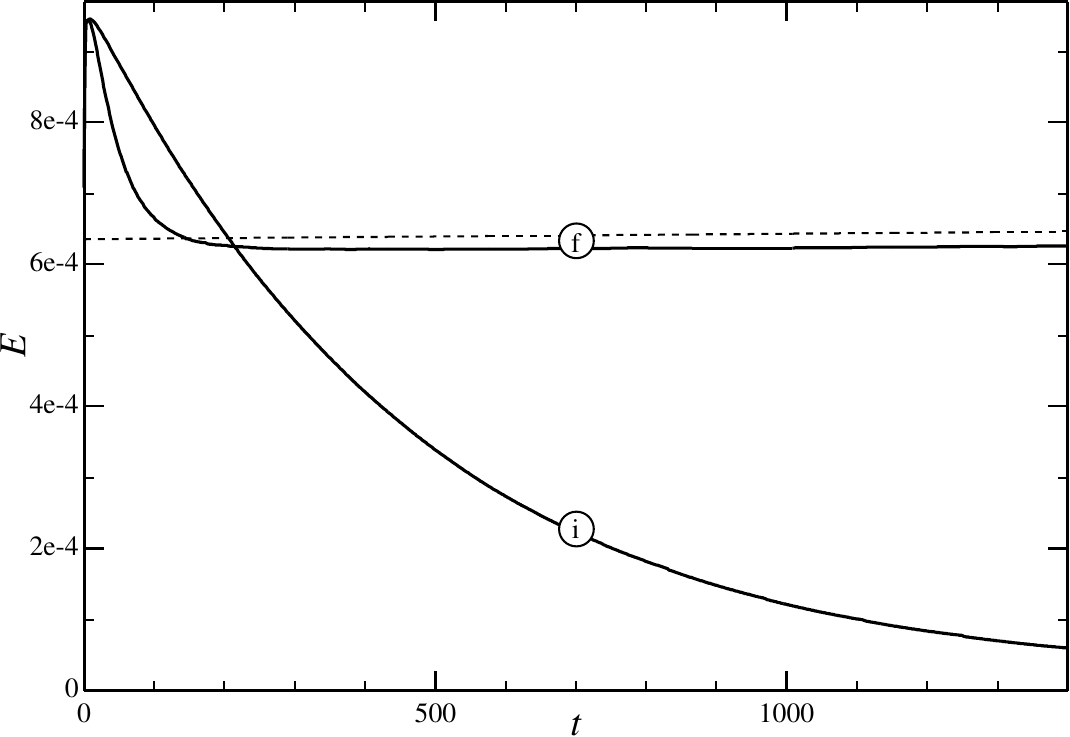}
\caption{The evaporation rate vs. time, for the computations represented in Fig. \ref{fig7}: (i) insulated substrate, (f) fixed-temperature substrate [the dotted curve shows the asymptotic solution (\ref{4.32} )--(\ref{4.33})].}
\label{fig8}
\end{figure}

Figs. \ref{fig7}--\ref{fig8} illustrate a typical evolution. It was computed
for the viscosity $\eta$ given by (\ref{3.9}) and%
\[
c_{V}=3,\qquad\kappa=\kappa_{0}\rho,
\]
where the former value corresponds to diluted water vapor, and the latter
reflects the general tendency of thermal conductivity to increase with
density. Comparing the nondimensionalization of this paper with that of Ref.
\cite{Benilov23a}, one can deduce that $\kappa_{0}$ is the reciprocal of the
\textquotedblleft isothermality parameter\textquotedblright\ $\beta$
introduced in the latter: if, in a certain region of the flow, $\kappa_{0}$ is
large ($\beta$, small), the temperature field in this region is almost
uniform. According to the estimate of Ref. \cite{Benilov23a} for water under
normal conditions, $\beta\approx0.06$, which approximately corresponds to%
\[
\kappa_{0}=17.
\]
The simplest initial condition for the temperature was used,
$T=\operatorname{const}$, and the initial density field was represented by the
steady isothermal solution -- i.e., that of (\ref{3.2}), (\ref{3.4}),
(\ref{3.6})--(\ref{3.7}) -- for $\rho^{(v)}/\rho^{(v.sat)}=0.9$, shifted to
the right by a distance of $\Delta z=20$ from the substrate. The boundary
conditions for the vapor density and temperature were moved from $z=+\infty$
to $z=400$, and the computation was stopped well before the perturbations of
$T$ and $\rho$ reached anywhere near this point.

The following features of the evolution can be observed:

\begin{itemize}
\item Fig. \ref{fig7} shows that the evolution of the density field is much
slower than the temperature evolution (as assumed in the asymptotic analysis).
This conclusion applies to both kinds of substrates.

\item The two lower left-hand panels of Fig. \ref{fig7} show that, for the
fixed-$T$ substrate, the temperature field in the liquid settles into a
quasi-steady pattern with linear dependence of $T$ on $z$ and uniform heat
flux [as predicted by asymptotics (\ref{4.17})]. Fig. \ref{fig8} shows that
the evaporation rate is changing very slowly in this case and is close to its
asymptotic value predicted by formulae (\ref{4.32})--(\ref{4.33}).

\item The two right-hand lower panels of Fig. \ref{fig7} show that, for an
insulated substrate, the liquid's temperature becomes nearly-uniform [as
predicted by asymptotics (\ref{4.18})] and is rapidly decreasing, while $E$
tends to zero (see Fig. \ref{fig8}). Since $T$ cannot drop below the dew point
(at which the initial density of vapor coincides with the saturated density),
it can only approach it from above\footnote{Calculating the dew point for the
parameters of Figs. \ref{fig7}--\ref{fig8}, one obtains $\left(
T-T_{0}\right)  \times10^{3}\approx-4.294$. The convergence to this value in
the actual figures is extremely slow, making it difficult to extend the
simulation to the stage where it is almost reached.}.

\item This effectively means that, in the case of insulated substrate, the
evaporation rate depends on $t$ -- hence, is not fully determined by a set of
external (measurable) parameters. It can only be measured directly.

\item For fixed-$T$ substrates, in turn, the evaporation rate does not tend to
zero and \emph{is} determined by external parameters (as evidenced by the good
agreement between the dotted and solid curves, both marked with (f), in Fig.
\ref{fig8}).
\end{itemize}

\subsection{Can the problem be solved if condition (\ref{4.10}) does not
hold?\label{Sec 4.4}}

If the depth $D$ of the evaporating layer is large enough, the timescale of
the long-term temperature evolution is comparable to that of evaporation. This
invalidates the asymptotic approach used in Sec. (\ref{Sec 4.1}) for solving
the macroscopic problem. Another shortcoming of the asymptotic approach is
that it works only if the relative humidity is close to unity.

However, even though there is no timescale separation in this case, one can
still exploit the \emph{spatial} scale separation -- i.e., the difference
between the microscopic scale of the interface and the macroscopic scale of
the liquid layer. This implies solving the macroscopic equations
(\ref{4.3})--(\ref{4.4}) numerically and, at each time step, feeding the
computed temperature of, and fluxes at, the interface into the microscopic
problem -- then using it to compute $E$ from the microscopic problem and
feeding $E$ back into the macroscopic equations.

The approach outlined above will not be discussed in further detail. It is
worth implementing only after the present model is extended to evaporation
into air (as opposed to the liquid's own vapor).

\section{Can other factors contribute to the discrepancies among experimental
results?\label{Sec 5}}

There are two factors, not included in the present model, which may
potentially contribute to the discord among the empiric formulae illustrated
in Fig. \ref{fig1}. These factors are convection in the liquid and diffusion
of the vapor in air.

\begin{itemize}
\item In a layer on a fixed-$T$ substrate, evaporation can cause a large
temperature difference between the substrate and interface -- which, in turn,
can trigger off convection. The size and number of the convection cells depend
on the depth $D$ of the tank with liquid, but also on its width $L$ -- hence,
two experiments differing by the value of $L$ but identical otherwise may
yield different results. It is not a priory clear how large this difference
is, but it is telling that some of the empiric formulae listed in Refs.
\cite{PoosVarju20,VarjuPoos22} do involve $L$.

\item It is well known that diffusion of vapor, or any other substance, in an
\emph{infinite} semispace does not have a steady regime: the one-dimensional
diffusion equation does not simply have such solutions. Physically, the vapor
accumulates near the interface in this case, making the evaporation rate tend
to zero with time -- while the layer of nearly-saturated vapor slowly expands
toward infinity.\\*\hspace*{0.5cm}If, however, the vapor is collected at a
\emph{finite} height above the interface, a steady regime does exist, such
that the vapor flux is spatially uniform and vapor concentration is linear.
This effectively means that the measured evaporation rate should depend on the
evaporation chamber's height. Such a dependence is indeed acknowledged in some
papers (e.g., Ref. \cite{SparrowKratzSchuerger83}), but not mentioned in the
others (including the results presented in Fig. \ref{fig1}).
\end{itemize}

Both of the above factors can be incorporated into the diffuse-interface
model: to study convection, a three-dimensional version of the DIM should be
used, whereas diffusion of vapor in air can be examined via a multicomponent
version \cite{Benilov23a}. The work on the latter problem is in progress.

Among other factors affecting evaporation, one might mention radiative
exchange of heat between liquid and air. It is often small, but can still be
important for the case of insulated vessels, where the heat fluxes from the
boundaries are zero.

\section{Summary and concluding remarks\label{Sec 6}}

The following features of nonisothermal evaporation of liquid layers on a
substrate are reported in this paper:

\begin{enumerate}
\item[(a)] Qualitative estimates (\ref{1.1}) and (\ref{1.5}) show that
nonisothermality affects evaporation even under normal conditions (and
probably more so in high-temperature high-pressure industrial processes).

\item[(b)] Calculations and simulations show that, if the substrate is
insulated, the temperature decreases toward the dew point, while the
evaporation rate $E$ tends to zero. This implies that $E$ cannot be deduced by
measuring the external parameters only.

\item[(c)] If the substrate is maintained at a fixed temperature, the heat
flux coming from the substrate supports evaporation at a finite rate. The heat
flux in the liquid is spatially uniform and the temperature profile is linear,
which agrees with measurements \cite{WardStanga01}. Asymptotic formula
(\ref{4.32}) has been obtained, relating $E$ to the fluid's parameters, the
layer's depth, and relative humidity (which was assumed to be close to unity).
\end{enumerate}

\noindent Another conclusion has been drawn for the limit of isothermal evaporation:

\begin{enumerate}
\item[(d)] If the temperature is sufficiently high and the vapor density is
sufficiently low, the vapor pressure cannot be matched by that of the liquid
(as illustrated in Fig. \ref{fig3}a). In such cases, the low pressure strains
liquid, encouraging cavitation -- and it can be conjectured that `near-vacuum
boiling' occurs.\newline\hspace*{0.5cm}If, however, the vapor pressure does
have a match for a liquid state, the pressure does equilibrate -- but the
chemical potentials of the two phases are still different. In this case,
evaporation occurs without boiling. The existence of such regimes on the
$(\rho^{(v)},T)$ plane is illustrated in Fig. \ref{fig3}b.
\end{enumerate}

\noindent Conclusions (b)--(d) have been drawn using the \emph{pure-fluid}
version of the diffuse-interface model, where air is approximated by vapor of
the same fluid. These results should rather be viewed as a proof of concept,
not an accurate predictive tool. To obtain the latter, one needs to use the
\emph{multicomponent} version of the DIM \cite{Benilov23a}.

There is one result, however, that can be extended to evaporation into air
with no further work -- namely, expressions (\ref{4.15}) for the heat fluxes
toward an interface in an unbounded space. Denoting these fluxes by
$Q^{(air)}$ and $Q^{(liquid)}$, one can deduce from (\ref{4.15}) that%
\begin{equation}
\frac{Q^{(air)}}{Q^{(liquid)}}=-\sqrt{\frac{\left(  \rho c_{p}\kappa\right)
^{(air)}}{\left(  \rho c_{p}\kappa\right)  ^{(liquid)}}},\label{6.1}%
\end{equation}
where the minus reflects the opposite directions of the fluxes. Using Refs.
\cite{TheEngineeringToolbox-AirDensity,TheEngineeringToolbox-AirSpecificHeat,TheEngineeringToolbox-AirThermalConductivity}
and \cite{LindstromMallard97} to estimate the numerator and denominator of the
fraction on the right-hand side of (\ref{6.1}), one obtains for water and air
at $25^{\circ}\mathrm{C}$%
\[
\left\vert \frac{Q^{(air)}}{Q^{(liquid)}}\right\vert \approx0.0035.
\]
Evidently, the heat flux coming from air should be accounted for only if it is
the \emph{only} heat flux (which it indeed is for liquids in an insulated vessel).

It is interesting to speculate how the results of this paper can be extended
to droplets.

Evaporation of a \emph{sessile} droplet is probably similar to that of a
liquid layer on a fixed-$T$ substrate (since the droplet is small, it cannot
change the substrate's temperature). A \emph{floating} droplet, however, is
likely to behave differently. Since the only source of vaporization heat in
this case is air (whose density and thermal conductivity are small),
evaporation should be slow. Evaporative cooling for such droplets should be as
important as for a liquid layer on an insulated substrate.

\acknowledgments{The author is grateful to Daniel Jakubczyk and Tibor Poós for elucidating discussions of experimental results on evaporation.}

\appendix

\section{The van der Waals force as described by the diffuse-interface
model\label{Appendix A}}

The DIM is based on the following assumptions:

\begin{enumerate}
\item[(i)] The long-range attractive intermolecular force (van der Waals
force) can be modelled by a pair-wise isotropic potential $\Phi(r)$ where $r$
is the distance between two molecules. The net force affecting a given
molecule is the algebraic sum of the forces exerted on it by the other molecules.

\item[(ii)] The spatial scale of the van der Waals force is much smaller than
the interfacial thickness.\smallskip
\end{enumerate}

Now, consider a three-dimensional fluid with molecular mass $m$, so that
$\rho/m$ is the number density. According to assumption (i), the density of
the collective force exerted by the molecules at a point $\mathbf{r}$, is%
\begin{equation}
\mathbf{F}(\mathbf{r},t)=\frac{\rho(\mathbf{r},t)}{m}\mathbf{\nabla}\int%
\frac{\rho(\mathbf{r}^{\prime},t)}{m}\Phi(|\mathbf{r}^{\prime}-\mathbf{r|}%
)\,\mathrm{d}^{3}\mathbf{r}^{\prime}.\label{A.1}%
\end{equation}
Then, according to assumption (ii), let the spatial scale of $\rho
(\mathbf{r},t)$ be much larger than that of $\Phi(r)$, in which case
expression (\ref{A.1}) can be simplified asymptotically. To do so, change in
it $\mathbf{r}^{\prime}\rightarrow\mathbf{r}^{\prime}+\mathbf{r}$ and then
expand $\rho_{j}(\mathbf{r}^{\prime}+\mathbf{r},t)$ about $\mathbf{r}^{\prime
}$, which yields%
\begin{multline*}
\mathbf{F}_{i}(\mathbf{r},t)=\frac{\rho(\mathbf{r},t)}{m}\mathbf{\nabla}%
\int\left[  \rho(\mathbf{r},t)+\mathbf{r}^{\prime}\cdot\mathbf{\nabla}%
\rho(\mathbf{r},t)\vphantom{\frac{1}{2}}\right.  \\
+\left.  \frac{1}{2}\mathbf{r}^{\prime}\mathbf{r}^{\prime}:\mathbf{\nabla
\nabla}\rho(\mathbf{r},t)+\cdots\right]  \frac{\Phi(r^{\prime})}{m}%
\mathrm{d}^{3}\mathbf{r}^{\prime}.
\end{multline*}
Since $\Phi(r^{\prime})$ is isotropic, the second integral in the above
expansion vanishes, and one obtains%
\begin{equation}
\mathbf{F}=\rho\left(  C\mathbf{\nabla}\rho+K\mathbf{\nabla}\nabla^{2}%
\rho+\cdots\right)  ,\label{A.2}%
\end{equation}
where%
\[
C=\int\frac{\Phi(r^{\prime})}{m^{2}}\mathrm{d}^{3}\mathbf{r}^{\prime},\qquad
K=\int\frac{r^{\prime2}}{2}\frac{\Phi(r^{\prime})}{m^{2}}\mathrm{d}%
^{3}\mathbf{r}^{\prime}.
\]
After the substitution of (\ref{A.2}) into the hydrodynamic equations, the
term involving $C$ can be absorbed into the internal energy -- i.e.,
eliminated by changing%
\[
e\rightarrow e+\frac{C}{2}\rho,\qquad p\rightarrow p+C\rho^{2}.
\]
This reflects the fact that the energy of molecular interactions is a kind of
internal energy.

Thus, without loss of generality, one can set in expression (\ref{A.2}) $C=0$.
Omitting also the small terms hidden in \textquotedblleft$\cdots
$\textquotedblright, one obtains%
\[
\mathbf{F}=K\rho\mathbf{\nabla}\nabla^{2}\rho.
\]
The one-dimensional reduction of this expression represents the van der Waals
force in Eqs. (\ref{2.10})--(\ref{2.11}).

\section{Properties of boundary-value problem (\ref{3.10})--(\ref{3.12}%
)\label{Appendix B}}

\subsection{The Maxwell construction (\ref{2.15})-- (\ref{2.16}%
)\label{Appendix B.1}}

To verify Eq. (\ref{2.15}), consider the limit $\xi\rightarrow-\infty$ in Eq.
(\ref{3.10}). Taking into account boundary condition (\ref{3.11}), one
immediately obtains (\ref{2.15}).

To verify Eq. (\ref{2.16}), differentiate (\ref{3.10}) and use thermodynamic
identity (\ref{2.4}) to obtain%
\[
\frac{\mathrm{d}^{3}\rho^{(0)}}{\mathrm{d}z^{3}}-\frac{\partial G(\rho
^{(0)},T)}{\partial\rho^{(0)}}\frac{\mathrm{d}\rho^{(0)}}{\mathrm{d}z}=0.
\]
Integrating this equality and fixing the constant of integration via boundary
condition (\ref{3.12}), one obtains%
\begin{equation}
\frac{\mathrm{d}^{2}\rho^{(0)}}{\mathrm{d}\xi^{2}}-G(\rho^{(0)},T)=-G(\rho
^{(v.sat)},T).\label{B.1}%
\end{equation}
Taking in the above equation the limit $\xi\rightarrow-\infty$ and recalling
boundary condition (\ref{3.11}), one recovers Eq. (\ref{2.16}).

\subsection{Derivation of formulae (\ref{3.20})--(\ref{3.21}%
)\label{Appendix B.2}}

First, observe that identity (\ref{2.4}) implies that%
\begin{equation}
G=\frac{\partial\left(  \rho G-p\right)  }{\partial\rho}.\label{B.2}%
\end{equation}
Now, multiply Eq. (\ref{B.1}) by $\mathrm{d}\rho^{(0)}/\mathrm{d}\xi$ and
integrate. Using identity (\ref{B.2}) and boundary condition (\ref{3.12}), one
obtains, after straightforward algebra,\begin{widetext}%
\begin{equation}
\frac{1}{2}\left(  \frac{\mathrm{d}\rho^{(0)}}{\mathrm{d}\xi}\right)
^{2}=\rho^{(0)}\left[  G(\rho^{(0)},T)-G(\rho^{(v.sat)},T)\right]
-p(\rho^{(0)},T)+p(\rho^{(v.sat)},T).\label{B.3}%
\end{equation}
Given that $\rho^{(0)}(\xi)$ is supposed to be a decreasing function, it follows
from (\ref{B.3}) that (the superscript $^{(0)}$ omitted)%
\[
\mathrm{d}\rho=-2^{1/2}\sqrt{\rho\left[  G(\rho,T)-G(\rho^{(v.sat)},T)\right]
-p(\rho,T)+p(\rho^{(v.sat)},T)}\,\mathrm{d}\xi.
\]
\end{widetext}This result allows one to transform (\ref{3.19}) into
(\ref{3.20}).

To obtain expression (\ref{3.21}), assume that $T$ is small -- hence,
$\rho^{(v.sat)}$ is also small, and the main contribution to integral
(\ref{3.20}) comes from the region near the lower limit of integration. Since
the density is small there, the general expressions for the chemical potential
and pressure can be replaced with their ideal-gas limits,%
\[
G(\rho,T)\sim T\ln\rho,\qquad p(\rho,T)\sim T\rho,
\]
and the viscosity can be replaced with its small-density limit,%
\[
\eta(\rho,T)\sim\eta(0,T).
\]
Since $\rho^{(l.sat)}\gg\rho^{(v.sat)}$, the upper limit of (\ref{3.20}) can
be replaced with $\infty$, and one obtains%
\[
A\sim\frac{2^{1/2}\eta(0,T)\,T^{1/2}}{\left(  \rho^{(v.sat)}\right)  ^{5/2}%
}\int_{1}^{\infty}\frac{\sqrt{\hat{\rho}\ln\hat{\rho}-\hat{\rho}+1}}{\hat
{\rho}^{4}}\mathrm{d}\hat{\rho},
\]
where $\hat{\rho}=\rho/\rho^{(v.sat)}$. The integral in this expression can be
evaluated numerically, yielding Eq. \ref{3.21}).

\section{The temperature equation\label{Appendix C}}

\subsection{Reduction of the energy equation to the temperature
equation\label{Appendix C.1}}

Replace $\partial e/\partial t$ in Eq. (\ref{2.29}) with%
\[
\frac{\partial e}{\partial t}=\frac{\partial e}{\partial\rho}\frac
{\partial\rho}{\partial t}+\frac{\partial e}{\partial T}\frac{\partial
T}{\partial t},
\]
and then eliminate $\partial\rho/\partial t$ using the density equation
(\ref{2.27}). Recalling definition (\ref{2.6}) of the heat capacity at
constant volume, one obtains%

\begin{multline}
\rho c_{V}\left(  \frac{\partial T}{\partial t}+w\frac{\partial T}{\partial
z}\right)  +B\frac{\partial w}{\partial z}=\eta\left(  \frac{\partial
w}{\partial z}\right)  ^{2}\\
+\frac{\partial}{\partial z}\left(  \kappa\frac{\partial T}{\partial
z}\right)  , \label{C.1}%
\end{multline}
where%
\begin{equation}
B=p-\rho^{2}\frac{\partial e}{\partial\rho} \label{C.2}%
\end{equation}
characterizes the production (consumption) of thermal energy due to mechanical
compression (expansion) of the fluid. The first term on the right-hand side of
(\ref{C.1}) describes the production of heat by viscosity.

Eq. (\ref{C.1}) is explicitly resolved with respect to $\partial T/\partial t$
and, thus, is more convenient for computations than Eq. (\ref{2.29}).

\subsection{The heat conduction equation\label{Appendix C.2}}

Recall ansatz (\ref{4.1}) where the temperature field was decomposed into a
background value $T_{0}$ and a small variation $\tilde{T}$. To use it in the
forthcoming formal derivation, introduce a small parameter $\epsilon$, and let%
\[
T=T_{0}+\epsilon\tilde{T}.
\]
A similar ansatz is applied to the density field,%
\[
\rho=\rho_{0}+\epsilon\tilde{\rho},
\]
the velocity will be scaled via%
\[
w=\epsilon^{1/2}\tilde{w},
\]
and the free variables, via%
\[
z=\epsilon^{-1/2}z_{new},\qquad t=\epsilon^{-1}t_{new}.
\]
Recalling that the original nodimensionalization in Sec. \ref{Sec 2.3} assumed
the spatial scale to be that of the interface, the stretched variable
$z_{new}$ implies that now one considers a region \emph{far above} or
\emph{far below} the interface.

Rewriting Eqs. (\ref{2.27})--(\ref{2.28}) and (\ref{C.1}) in terms of the new
variables, omitting the subscript $_{new}$, and keeping the leading-order
only, one obtains%
\begin{equation}
\frac{\partial\tilde{\rho}}{\partial t}+\rho_{0}\frac{\partial w}{\partial
z}=0,\label{C.3}%
\end{equation}%
\begin{equation}
\frac{\partial}{\partial z}\left[  \frac{\partial p(\rho_{0},T_{0})}%
{\partial\rho_{0}}\tilde{\rho}+\frac{\partial p(\rho_{0},T_{0})}{\partial
T_{0}}\tilde{T}\right]  =0,\label{C.4}%
\end{equation}%
\begin{equation}
\rho_{0}c_{V}(\rho_{0},T_{0})\frac{\partial\tilde{T}}{\partial t}+B(\rho
_{0},T_{0})\frac{\partial w}{\partial z}=\kappa(\rho_{0},T_{0})\frac
{\partial^{2}\tilde{T}}{\partial z^{2}}.\label{C.5}%
\end{equation}
The parameter $\epsilon$ has now played its role of an `indicator' of small
terms, as they have all been omitted. One can therefore set $\epsilon=1$, so
that the rescaled variables coincide with those used in the main body of the paper.

Next, integration of Eq. (\ref{C.4}) plus an assumption that the constant of
integration is zero (which amounts to the requirement that the pressure at
infinity does not vary in time) yield%
\[
\tilde{\rho}=-\frac{\partial p}{\partial T}\left[  \frac{\partial p(\rho
_{0},T_{0})}{\partial\rho_{0}}\right]  ^{-1}\tilde{T}.
\]
Using this equality and Eq. (\ref{C.3}), one obtains%
\[
\frac{\partial w}{\partial z}=\frac{1}{\rho_{0}}\frac{\partial p(\rho
_{0},T_{0})}{\partial T_{0}}\left[  \frac{\partial p(\rho_{0},T_{0})}%
{\partial\rho_{0}}\right]  ^{-1}\frac{\partial\tilde{T}}{\partial t},
\]
so that Eq. (\ref{C.5}) becomes (the subscript $_{0}$ omitted)%
\begin{equation}
\left[  \rho c_{V}+\frac{B}{\rho}\frac{\partial p}{\partial T}\left(
\frac{\partial p}{\partial\rho}\right)  ^{-1}\right]  \frac{\partial\tilde{T}%
}{\partial t}=\kappa\frac{\partial^{2}\tilde{T}}{\partial z^{2}}.\label{C.6}%
\end{equation}
Finally, definition (\ref{C.2}) of $B$ and definition (\ref{2.8}) of $c_{P}$
help one to reduce (\ref{C.6}) to the standard heat conduction equations, as required.

\section{Proof of identity (\ref{4.29})\label{Appendix D}}

By definition, the vaporization heat is%
\begin{equation}
\Delta h(T)=h(\rho^{(v.sat)},T)-h(\rho^{(l.sat)},T), \label{D.1}%
\end{equation}
where $\rho^{(v.sat)}$ and $\rho^{(l.sat)}$ are the densities of the saturated
vapor and liquid, respectively, and the enthalpy $h$ is given by (\ref{2.7}).
Recalling definition (\ref{2.3}) of the chemical potential (which implies that
$h=G+Ts$), one can transform (\ref{D.1}) into%
\begin{multline*}
\Delta h(T)=G(\rho^{(v.sat)},T)+Ts(\rho^{(v.sat)},T)\\
-G(\rho^{(l.sat)},T)-Ts(\rho^{(l.sat)},T).
\end{multline*}
Recalling equality (\ref{2.16}) (the second part of the Maxwell construction),
one obtains (\ref{4.29}), as required.

% \bibliography{../../bib/refs}
\bibliography{}

\end{document}